\documentclass[10pt,letterpaper]{article}
\usepackage[top=0.85in,left=2.75in,footskip=0.75in,marginparwidth=2in]{geometry}

\usepackage{nameref,hyperref}
\usepackage[right]{lineno}

\usepackage{microtype}

\raggedright
\usepackage{changepage}

\usepackage[aboveskip=1pt,labelfont=bf,labelsep=period,singlelinecheck=off]{caption}

\makeatletter
\renewcommand{\@biblabel}[1]{\quad#1.}
\makeatother

\usepackage{lastpage,fancyhdr,graphicx}
\fancyheadoffset[L]{2.25in}
\fancyfootoffset[L]{2.25in}

\usepackage{color}

\definecolor{Gray}{gray}{.25}

\usepackage{graphicx}

\usepackage{sidecap}

\usepackage{wrapfig}
\usepackage[pscoord]{eso-pic}
\usepackage[fulladjust]{marginnote}
\reversemarginpar

\usepackage{amsmath} \usepackage{amsfonts} \usepackage{amssymb} \usepackage{amsthm}
\usepackage{graphicx} \usepackage{color}
\usepackage{cases}
\usepackage{siunitx}
\usepackage[numbers]{natbib}
\usepackage{multirow}
\usepackage[table,xcdraw]{xcolor}
\usepackage{hyperref}

\begin{document}
\vspace*{0.35in}

\begin{flushleft}

{\Large
\textbf
\newline{Optimal signal transmission and timescale diversity in a model of human brain operating near criticality}
} \newline
\\
Yang Qi\textsuperscript{1,2,3,$\dagger$},
Jiexiang Wang\textsuperscript{1,$\dagger$},
Weiyang Ding\textsuperscript{1,2,3},
Gustavo Deco\textsuperscript{4,5},
Viktor Jirsa\textsuperscript{6},
Wenlian Lu\textsuperscript{7,1,2},
Jianfeng Feng\textsuperscript{1,2,3,$\ast$}
\\
\bigskip
{\bf 1} Institute of Science and Technology for Brain-Inspired Intelligence, Fudan University, Shanghai, China
\\
{\bf 2} Key Laboratory of Computational Neuroscience and Brain-Inspired Intelligence (Fudan University), Ministry of Education, China
\\
{\bf 3} MOE Frontiers Center for Brain Science, Fudan University, Shanghai, China 
\\
{\bf 4} Institució Catalana de Recerca i Estudis Avançats (ICREA), Barcelona, Spain.
\\
{\bf 5} Computational Neuroscience Group, Center for Brain and Cognition, Department of Information and Communication Technologies, Universitat Pompeu Fabra, Barcelona, Spain
\\
{\bf 6} Aix Marseille University INSERM, INS, Institute for Systems Neuroscience, 13005 Marseille, France
\\
{\bf 7} Center for Applied Mathematics, Fudan University, Shanghai, China
\\
\bigskip
$\ast$ jffeng@fudan.edu.cn

$\dagger$ These authors contributed equally

\end{flushleft}

\section*{Abstract}
Cortical neurons exhibit a hierarchy of timescales across brain regions in response to input stimuli, which is thought to be crucial for information processing of different temporal scales. Modeling studies suggest that both intra-regional circuit dynamics as well as cross-regional connectome may contribute to this timescale diversity. Equally important to diverse timescales is the ability to transmit sensory signals reliably across the whole brain. Therefore, the brain must be able to generate diverse timescales while simultaneously minimizing signal attenuation. To understand the dynamical mechanism behind these phenomena, we develop a second-order mean field model of the human brain by applying moment closure and coarse-graining to a digital twin brain model endowed with whole brain structural connectome. Cross-regional coupling strength is found to induced a phase transition from asynchronous activity to synchronous oscillation. By analyzing the input-response properties of the model, we reveal criticality as a unifying mechanism for enabling simultaneously optimal signal transmission and timescales diversity. We show how structural connectome and criticality jointly shape intrinsic timescale hierarchy across the brain.


\section{Introduction}

The human brain is a highly heterogeneous and complex system, with distinct regions serving specialized functions. The coordinated interactions among these regions collectively shape a hierarchical mode of information processing. Recent experimental evidence suggests that neural response to input stimuli exhibits a timescale gradient across the whole brain in both human and non-human primates~\cite{Wolff2022trends,Golesorkhi2021commbio,Senkowski2024nrn,Cavanagh2020frontiers,Honey2012neuron}. In particular, sensory areas respond with brief, transient reactions to input stimuli whereas higher-order cortical areas produce responses that remain persistent over longer durations~\cite{Chang2022pnas,Honey2012neuron,Himberger2018neurosci}. This hierarchy of intrinsic timescales are thought to play key roles in processing sensory inputs and generating behavior at different temporal scales~\cite{Golesorkhi2021commbio,Chang2022pnas,Cavanagh2016elife,Zeraati2023natcomms}.

Modeling studies have attributed the emergence of timescale hierarchy to heterogeneity in self-coupling~\cite{Stern2023elife}, local recurrent connectivity~\cite{Chaudhuri2014eLife,Huang2017curr_op_neurobio}, and whole brain anatomical connectome~\cite{Li2022pnas,chaudhuri2015large}. Notably, computational modeling of the monkey brain has revealed that the intrinsic timescales gradually increase from area to area along the cortical hierarchy~\cite{Li2022pnas,chaudhuri2015large}. 
Using linear network models, it has been shown that intrinsic timescales become segregated if the eigenmodes of the connectivity matrix are localized to different parts of the network~\cite{Chaudhuri2014eLife,Li2022pnas}. Such localization of eigenmodes may arise from heterogeneity in spatially local connectivity~\cite{Chaudhuri2014eLife} as well as connectivity endowed with anatomical connectome~\cite{Li2022pnas}. 
Despite these theoretical insights, how intrinsic neural timescales may arise in the human brain has yet been explored.

Although previous modeling studies have focused on explaining the origin of timescale hierarchies in the brain, the signal transmission aspect of input processing has not been fully addressed. If input signals attenuate too quickly as it propagates across the cortex, then responses in higher order cortical areas with long timescales may become negligible. This raises the question how the brain might attain diverse timescales while simultaneously ensuring reliable signal transmission. In nature, many physical systems operate in a regime that exhibits signatures of criticality, which is thought to be optimal for signal transmission~\cite{Haken1977,Meijers2021pre}. 
This has inspired the critical brain hypothesis which suggests that the brain may gain significant information-processing advantages by operating near the critical point of a phase transition, including heightened sensitivity to perturbations, a diverse range of system states, maximal brain fluidity, and increased capacity for information storage and transmission~\cite{o2022critical,munoz2018colloquium,Cocchi2017progress,Spiegler2016eneuro,Breyton2024elife,Lavanga2023neuroimage,Huang2017curr_op_neurobio}.

In this study, we develop a second-order dynamic mean field model of the human brain by applying moment closure and coarse-graining to a recently proposed Digital Twin Brain (DTB) model for simulating the whole brain at single neuron resolution~\cite{Lu2024nsr,Lu2024nat_cs}. By globally tuning the strength of synaptic connections between brain regions, we show that the mean field model exhibits a phase transition from asynchronous irregular activity to brain-wide synchronous oscillations, as consistent with the DTB model. Using the mean field model, we investigate the processing of visual sensory signals across the whole brain, with special focuses on the strength of signal transmitted to each brain region and the hierarchy of intrinsic timescales as the system state approaches criticality.

\section{Results}

\subsection{Moment closure, coarse-graining, and a mean field model for the human brain}
\label{Regionwise_MNNmodel}

To link whole brain dynamics at the macroscale to biophysical nonlinearity at the microscale, we develop a multiscale approach to whole brain modeling, as illustrated schematically in Fig.~\ref{fig:fig2}(A). 

We begin with a region-wise computational cortico-subcortical model of whole human brain at single neuron resolution~\cite{Lu2024nsr,Lu2024nat_cs}. The model comprises $n=378$ cortical and subcortical regions according to the brain parcellation HCPex~\cite{huang2022extended}. Each area is represented by a conductance-based spiking neural circuit containing one excitatory (AMPA-ergic) population and one inhibitory (GABA-ergic) population. Different brain regions are connected only by synaptic projections between excitatory neurons. 
Each neuron is modeled using a conductance-based leaky integrate-and-fire neuron with finite synaptic time constants. The synaptic connectivity within and between regions are drawn randomly according to certain distributions with given average in-degrees. The intra-regional and the cross-regional synaptic in-degrees, the conductance of each region, and the number of neurons within each brain region are estimated based on biological constraints (see Methods). External inputs to different brain regions are determined via a data assimilation approach~\cite{Lu2024nsr,Lu2024nat_cs}. A global scaling factor $\gamma$ controls the cross-regional synaptic coupling strength.

We apply moment closure and coarse-graining to arrive at a second-order mean field model for the human brain. The moment closure is performed on the spiking neural circuit model through a Fokker-Planck formalism. The coarse-graining is performed by treating each brain region as a homogeneous population of neurons with identical firing statistics. Under these treatments, the state vector $\mathbf{m}$ of the mean field model is $\mathbf{m}=(\boldsymbol{\mu}_E,\boldsymbol{\sigma}^2_E,\boldsymbol{\mu}_I,\boldsymbol{\sigma}^2_I)\in\mathbb{R}^{4n}$, where $n=378$ is the number of brain regions. Here, $\boldsymbol{\mu}_\alpha=(\mu^\alpha_u)$ and $\boldsymbol{\sigma}^2_\alpha=(\sigma^{\alpha,2}_u)$ represent the mean firing rate and firing variability of neural population $\alpha\in\{E,I\}$ in region $u$. We can then write down the mean field model as
\begin{equation}
    \begin{aligned}
\tau\dfrac{d}{dt} \left(
\begin{array}{c}
\boldsymbol{\mu}_E \\
\boldsymbol{\sigma}^2_E\\\boldsymbol{\mu}_I \\
\boldsymbol{\sigma}^2_I\\
\end{array} \right)  = - \left(
\begin{array}{c}
\boldsymbol{\mu}_E \\
\boldsymbol{\sigma}^2_E\\\boldsymbol{\mu}_I \\
\boldsymbol{\sigma}^2_I\\
\end{array} \right) + \Phi\left(
\begin{array}{c}
\boldsymbol{\mu}_E \\
\boldsymbol{\sigma}^2_E\\\boldsymbol{\mu}_I \\
\boldsymbol{\sigma}^2_I\\
\end{array} \right). \label{eq:self-consistent equations regarding firing statistics}
    \end{aligned}
\end{equation}
The time constant $\tau$ is determined to be approximately 11 ms by fitting the oscillation frequency of the mean field model to that of the spiking model. The coarse-grained moment mapping $\Phi$ can be decomposed into
\begin{equation}
    \Phi = \phi_{\rm{MA}} \circ \phi_{\rm{eff}} \circ \phi_{\rm{sum}}.
    \label{eq:Phi_decomposition}
\end{equation} 
The first component is the coarse-grained synaptic summation $\phi_{\rm sum}: \mathbb{R}^{4n}\to \mathbb{R}^{8n}$ which maps the moments of pre-synaptic spike count to that of synaptic activations at post-synaptic neurons. The second component is the effective current $\phi_{\rm eff}:\mathbb{R}^{8n}\to\mathbb{R}^{6n}$, whose output includes an effective time constant and the moments of the effective current. The third component $\phi_{\rm MA}: \mathbb{R}^{6n}\to\mathbb{R}^{4n}$ is the moment activation providing the moments of the output spike count. See Methods for details of these moment mappings.

 \subsection{Phase transition induced by cross-regional coupling}
\label{sec:mean field model captures macroscopic properties of the DTB model}

\begin{figure}
    \centering
    \includegraphics[width=\textwidth]{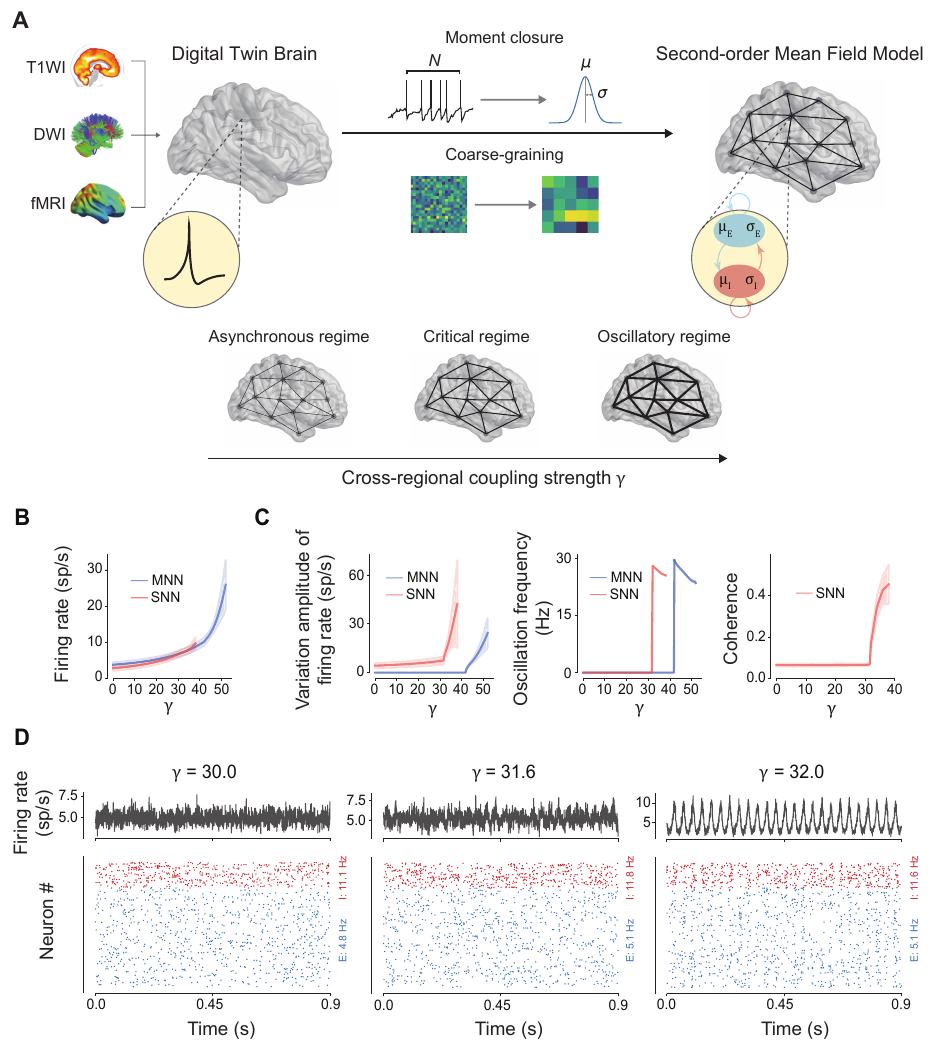}
\caption{\textbf{Phase transition induced by cross-regional coupling}. 
    (A) Construction of the digital twin brain (DTB) model for the whole human brain and a second-order mean field model through moment closure and coarse-graining. 
    (B) The mean firing rate of the mean field model closely matches that of the DTB model. Shades represent the standard deviation across brain regions. (C) Variation amplitude of population firing rate (left panel), oscillation frequency (middle panel), and membrane potential coherence (right panel) as a function of cross-regional coupling strength $\gamma$ reveal a phase transition from asynchronous activity to synchronous oscillatory activity in both the DTB model and the mean field model, with critical points at $\gamma=31.6$ and $\gamma = 41.8$, respectively. (D) Typical firing patterns in V1 of the DTB model for different dynamic regimes. Left: asynchronous activity ($\gamma = 30$); middle: near-critical ($\gamma = 31.6$); right: synchronous oscillatory activity ($\gamma = 32$). Lower panels show spike raster plots of 200 excitatory (blue) and 50 inhibitory (red) neurons from a representative brain region (V1). Upper panels show the population firing rate of all neurons in V1.
    }\label{fig:fig2}
\end{figure}

We investigate the dynamics of the DTB model and the mean field model by varying the global factor $\gamma$ for the cross-regional coupling strength. We find that as $\gamma$ increases, the mean firing rate averaged across all regions increases, with values obtained from the mean field model closely matching the DTB model (Fig.~\ref{fig:fig2}B). Simulations of the DTB reveal a phase transition from asynchronous activity to oscillatory activity as $\gamma$ increases. Specifically, when $\gamma$ is below a critical point $\gamma_c=31.6$, the DTB model is in a state of asynchronous activity with low firing rate ($<7$ sp/s) (Fig.~\ref{fig:fig2}B) and low fluctuation amplitude ($<8$ sp/s) (Fig.~\ref{fig:fig2}C, left panel). Above this critical point, the spiking neural circuit transitions into a state of synchronous oscillation with elevated firing rate (Fig.~\ref{fig:fig2}B) and a steep increase in the oscillation amplitude (Fig.~\ref{fig:fig2}C, left panel). Simulations of the mean field model show qualitatively consistent results with the critical point slightly shifted to $\gamma_c=41.8$. This mismatch is expected and is due to the analytical approximations in deriving the mean field model. In the synchronous regime, the oscillation frequencies of both the DTB and the mean field model decrease with $\gamma$ passing their respective critical points (Fig.~\ref{fig:fig2}C, middle panel). In addition, the phase transition is also reflected in the coherence $\rho$ in spiking activity within each brain region (Fig.~\ref{fig:fig2}C, right panel), defined as
$
    \rho \equiv \left(\frac{\sigma_V^2}{\sum_{i=1}^{N} \sigma_i^2 / {N}}\right)^{1/2},
$
where $\sigma_V^2$ represents the temporal variance of population-averaged membrane potential, $\sigma_i^2$ is the temporal variance of the membrane potential of individual neurons, and $N$ denotes the number of neurons within each brain region~\cite{Golomb1994physicad}. Perfect synchrony is indicated by $\rho=1$ and reduced synchrony corresponds to $\rho<1$. Spike raster plots and population firing rates for representative values of $\gamma$ for different regimes are shown in Fig.~\ref{fig:fig2}(D).

\subsection{Linear stability analysis and whole brain eigenmodes}
\label{sec:linear stability analysis}

\begin{figure}
    \centering
    \includegraphics[width=\textwidth]{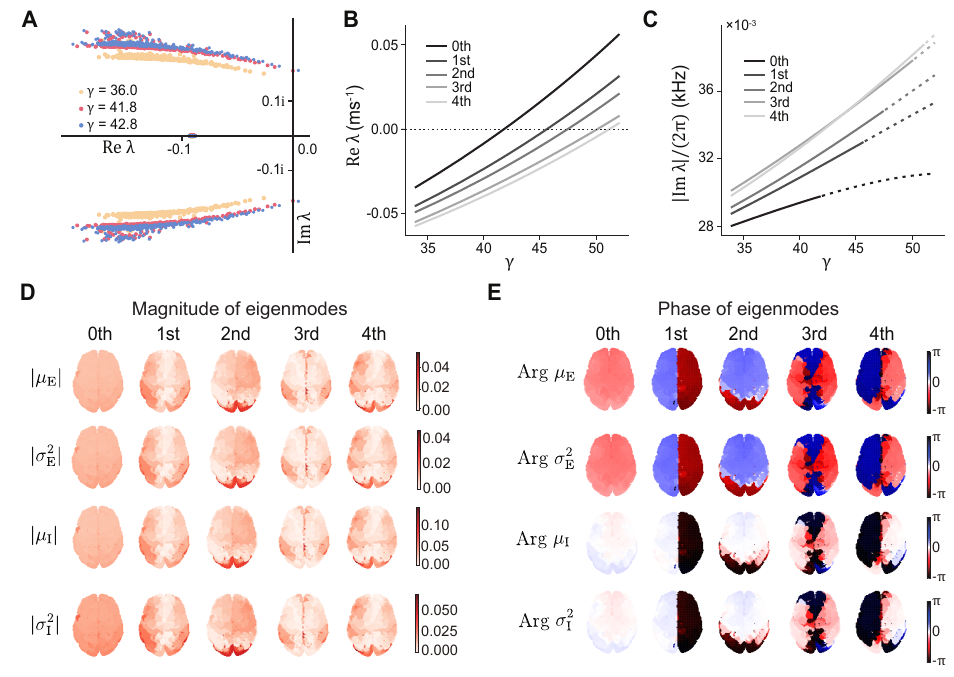}
    \caption{\textbf{Linear stability analysis and whole brain eigenmodes}. (A) Scatter plot of the eigenvalues on the complex plane for subcritical ($\gamma=36.0$), critical ($\gamma=41.8$), and supercritical ($\gamma=42.8$) regimes. (B) Real parts of the first few eigenvalues of the Jacobian matrix of the mean field model as a function of cross-regional coupling strength $\gamma$. (C) Oscillation frequencies of the first few eigenmodes as a function of $\gamma$. Solid and dotted lines correspond to stable and unstable eigenmodes, respectively. (D,E) Dorsal view of the magnitudes and phases of the first few eigenmodes at their respective critical points, revealing a hierarchical organization of spatial modes of increasing complexity.
    }
    \label{fig:Linear_stability_JFF_withsubcortical}
\end{figure}

We reveal the dynamical mechanism of the phase transition by performing a linear stability analysis on the mean field model. To this end, we linearize the system around its fixed points and then numerically calculate the eigenvalues and eigenmodes of the Jacobian matrix (see Methods). Figure \ref{fig:Linear_stability_JFF_withsubcortical}(A) shows the eigenvalues on the complex plane for representative values of cross-regional coupling strength $\gamma$. We find that when $\gamma$ is weak all eigenvalues have negative real parts, indicating that the system is stable. As $\gamma$ increases and crosses a critical point at $\gamma_c=41.8$, the eigenvalue with the largest real part crosses the imaginary axis. This indicates the emergence of an unstable eigenmode with an oscillation frequency of $f=|{\rm Im} \lambda| / (2\pi)$, which is found to be approximately $f=29.75$ Hz. As $\gamma$ further increases, more unstable eigenmodes with different frequencies emerge. We label the critical points at which the $l$-th unstable eigenmode emerges as $\gamma_c^{(l)}$.

The real and imaginary parts of eigenvalues corresponding to the first few eigenmodes are shown in Fig.~\ref{fig:Linear_stability_JFF_withsubcortical}(B) and (C), respectively. The magnitudes and phases of the eigenmodes as their corresponding eigenvalues cross the imaginary axis at critical points $\gamma_c^{(l)}$ are shown in Fig.~\ref{fig:Linear_stability_JFF_withsubcortical}(D)-(E). We find that the eigenmodes exhibit a hierarchy of spatial organization of increasing complexity. 
For all eigenmodes shown, no obvious difference is found between the mean firing rate and firing variability, whereas excitatory and inhibitory populations exhibit phase differences of around $\pi/4$. The eigenmodes derived from our mean field model is consistent with those found in linear systems embedded with structural connectome~\cite{Wang2019prl}.

\subsection{Enhancement of visual signal propagation near criticality}
\label{sec:Enhancement of visual signal propagation near criticality}

\begin{figure}
    \centering\includegraphics[width=\textwidth]{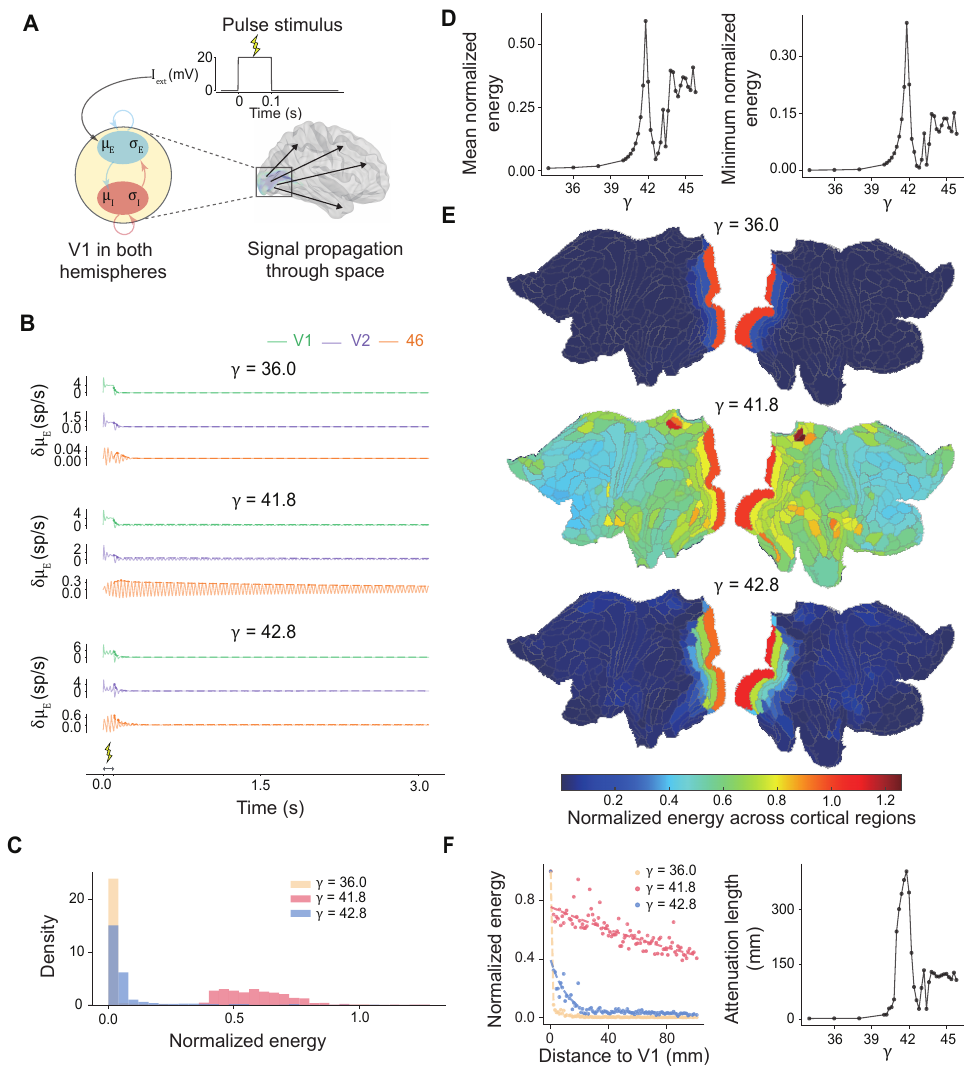}
    \caption{\textbf{Enhancement of visual signal propagation near criticality.}
(A) Schematics of the numerical experiments on signal dissipation during visual input processing. A 100 ms, 20 mV pulse stimulus is applied to V1 in both hemispheres. The model is simulated using a first-order Euler method with a time step of 0.001 ms. 
(B) Responses of representative regions (V1, V2, and area 46) for subcritical ($\gamma=36$), near-critical ($\gamma=41.8$), and supercritical ($\gamma=42.8$) regimes. (C) Distribution of response energy, normalized by the average energy of V1, across brain regions under different dynamical regimes. (D) Mean and minimum normalized response energy across regions for varying cross-regional coupling strengths $\gamma$. (E) Spatial organization of the normalized response energy across the cortical surface under different dynamical regimes. (F) The strength of visual signal as a function of Euclidean distance from the stimulation site (V1) (left panel). Dots, average normalized energy binned by distance; solid line, exponential fit. The attenuation lengths for varying cross-regional coupling strengths $\gamma$ is maximized near criticality (right panel).
\label{fig:Enhancement_propagation_power}}
\end{figure}

We test the effect of criticality on the dissipation of visual signals across brain regions in different dynamical regimes using the mean field model. This is done by numerical simulations of the model with a transient step input current added to visual region V1, mimicking stimulus presentation in biological experiments, as illustrated in Fig.~\ref{fig:Enhancement_propagation_power}(A) (see Methods). 
Figure \ref{fig:Enhancement_propagation_power}(B) presents the time series of mean firing rate response of excitatory populations in representative brain regions (V1, V2, and area 46) following current stimulation under subcritical ($\gamma=36.0$), critical ($\gamma=41.8$), and supercritical ($\gamma=42.8$) conditions. 
To quantify the sensitivity of each brain region to visual stimuli, we calculate the total energy of the response variations in mean firing rate of excitatory populations of each region as $E = \int_{t_0}^{\infty} |
\delta \mu_E(t)|^2dt$, where $t_0=100$ ms is the stimulus offset time. Higher total energy indicates stronger responses or greater sensitivity to sensory inputs, while lower energy reflects a weaker response or greater signal attenuation. This metric reflects both the peak amplitude and the persistence of the responses. 

In the subcritical and supercritical regimes, the relaxation time (time taken for the response envelope to decay from peak to half-maximum) is consistently of the same order of magnitude (as in Fig.~\ref{fig:Enhancement_propagation_power}(B)). However, the peak response amplitudes decrease substantially from V1 to V2 and further to area 46, showing notable differences in magnitude. Consequently, the total energy dissipates rapidly across regions, from V1 through V2 to area 46. In contrast, under the critical regime, while the peak amplitudes still decrease from V1 to V2 and area 46, area 46 exhibits a more persistent response compared to V1 and V2. As a result, under the critical regime, the dissipation of response energy across regions from V1 through V2 to area 46 is less pronounced compared to the subcritical and supercritical regimes.

We then analyze the distribution of energy across brain regions, normalized by the energy at the stimulation site (V1), as shown in Fig.~\ref{fig:Enhancement_propagation_power}(C). 
For the subcritical ($\gamma=36.0$) and supercritical ($\gamma=42.8$) conditions, the distribution of normalized energy is concentrated near 0. In contrast, near the critical regime ($\gamma=41.8$), the total energy exhibits a broad distribution, with values more comparable to V1. We also calculate the mean and minimum of normalized energy across all regions for varying $\gamma$ [Fig.~\ref{fig:Enhancement_propagation_power}(D)] and find that both metrics are peaked near the critical point, highlighting optimal signal propagation near criticality.

To understand the spatial aspects of signal transmission across brain regions under different dynamical regimes, we map the normalized energy on to the cortical surface, as shown in Fig.~\ref{fig:Enhancement_propagation_power}(E). In both subcritical and supercritical states, a sharp attenuation of energy is observed as signal propagates from V1 to the surrounding regions. In contrast, near the critical point, the spatial organization of energy exhibits a smoother gradient with much less signal attenuation. 
We further summarize the normalized energy in each region as a function of its Euclidean distance from the stimulation site (V1) [Fig.~\ref{fig:Enhancement_propagation_power}(F), left panel] and find that in both subcritical and supercritical regimes, the energy transmitted to different brain regions decays rapidly with distance from V1. However, in the critical regime, the energy remains persistent across all distances. The decay of energy with distance from V1 is fitted using an exponential function, $E(d)=A e^{-d/\Delta}$, where $\Delta$ defines the attenuation length. The right panel of Fig.~\ref{fig:Enhancement_propagation_power}(F) reveals that the attenuation length for varying $\gamma$ peaks near the critical point. Such divergence of attenuation length due to criticality is consistent with critical phenomena in physical systems~\cite{Haken1977,Meijers2021pre} including neural systems~\cite{Rabuffo2021eneuro,Spiegler2016eneuro,Lavanga2023neuroimage,Breyton2024elife}. These findings indicate that the spatial propagation of neural response signals experiences minimal attenuation near criticality.

\subsection{Optimal diversity of intrinsic timescales near criticality}
\label{sec:Nonlinear response properties of stimulus-evoked activity}

\begin{figure}
    \centering
\includegraphics[width=\textwidth]{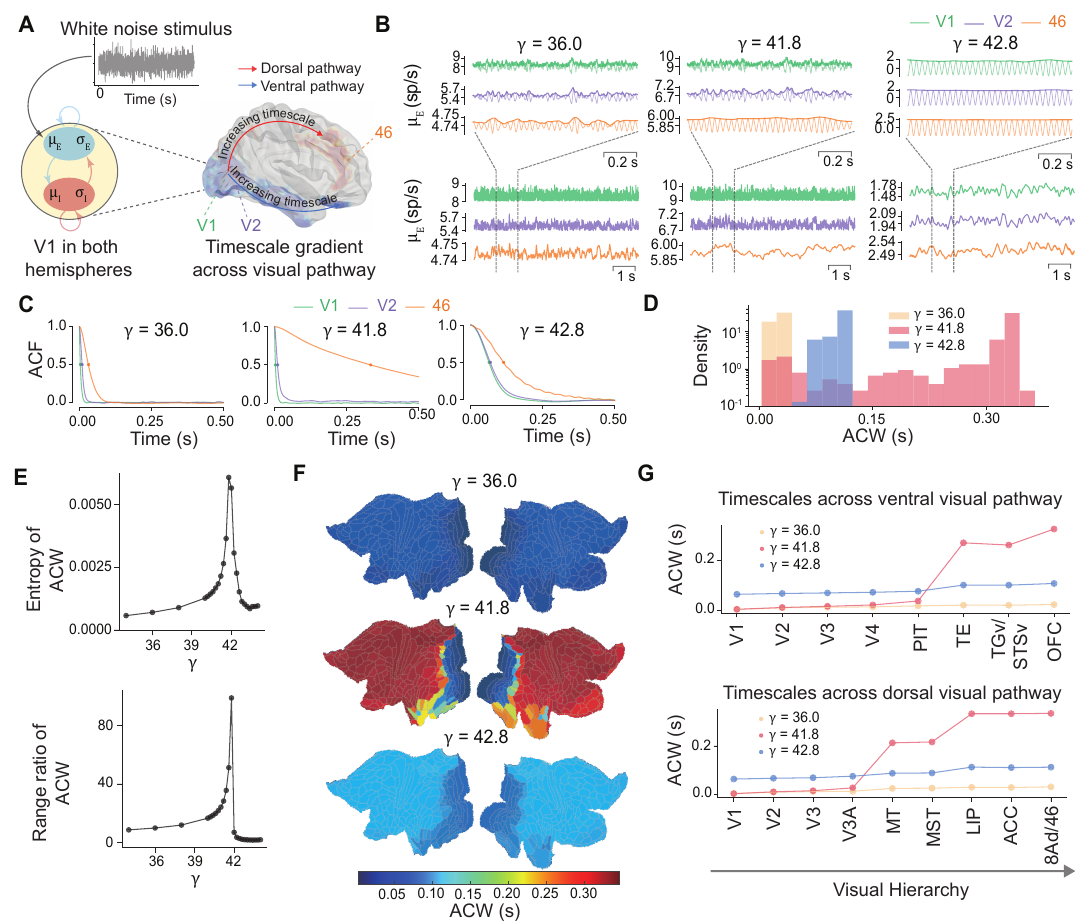}
    \caption{\textbf{Optimal diversity of intrinsic timescales near criticality.} (A) Schematics of the numerical experiment for estimating intrinsic timescales during visual signal propagation. The input signal, applied to V1 in both hemispheres, is modeled as Gaussian white noise with a mean of 0 mV and a standard deviation of 2 mV, whose autocorrelation function (ACF) approximates a Dirac delta function. The model is simulated using a first-order Euler method with a time step of 0.01 ms. (B) Responses in mean firing rate of excitatory populations of representative regions and their envelope. (C) The ACF of response envelopes for these regions. (D) Distribution of autocorrelation window (ACW) across brain regions under different dynamical regimes. (E) Entropy and range ratio of ACW as a function of cross-regional coupling strengths $\gamma$. (F) The ACW mapped to the cortical surface exhibiting a smoother spatial gradient near criticality. (G) Intrinsic timescales during visual signal processing along the ventral and dorsal visual pathways. 
    PIT, the posterior inferotemporal cortex; TE, anterior inferotemporal cortex; TGv, ventral temporal pole; STSv, inferior parts of the superior temporal sulcus; OFC, orbitofrontal cortex; MT, middle temporal area; MST, medial superior temporal area; LIP, lateral intraparietal area; ACC, anterior cingulate cortex. Results in panel (B)-(C) are obtained from a single trial; those in panel (D)-(G) are averaged over 100 trials.
    \label{fig:timescale}} \end{figure}

Beyond signal transmission strength, the diversity of timescales across brain regions represents another key characteristics of input processing in the brain. 
As shown in Fig.~\ref{fig:Enhancement_propagation_power}(B), when a brief stimulus is applied to V1, early sensory areas such as V1 and V2 exhibit brief, transient activity, while higher cortical regions such as area 46 generate responses that remain persistent over extended periods of time. However, the presence of both rise and decay times complicates the estimation of timescales. Therefore, we adopt the definition of intrinsic timescales as the temporal windows during which neural activity fluctuations remain strongly self-correlated, referred to as the autocorrelation window (ACW)~\cite{Murray2014nn,Li2022pnas}. Figure \ref{fig:timescale}(A) illustrates the experimental procedure for estimating ACW using continuous stimuli applied to V1 of both hemispheres. The input signal is modeled as a Gaussian white noise whose autocorrelation function approximates a Dirac delta function.

Figure \ref{fig:timescale}(B) presents the responses of representative brain regions (V1, V2, and area 46) and their signal envelopes for subcritical, critical, and supercritical regimes. In both subcritical and supercritical dynamical regimes, the response envelopes of these regions fluctuate on similarly short timescales. In contrast, under the critical regime, V1 and V2 continue to fluctuate rapidly while area 46's response envelope fluctuates more slowly over time. We calculate the autocorrelation function (ACF) of the response envelope and measure ACW as its half-life~\cite{Wolff2022trends,Golesorkhi2021commbio}. Figure \ref{fig:timescale}(C) displays the ACF for representative brain regions across different dynamical regimes. In both the subcritical and supercritical regimes, the activity of all displayed regions shows rapid decay in ACF. In contrast, under the critical regime, the activity of the high-order region (area 46) exhibits correlation over a longer period of time than the sensory regions (V1 and V2).

Next, we compare the distribution of intrinsic timescales across brain regions under different dynamical regimes, as illustrated in Fig.~\ref{fig:timescale}(D). In the subcritical and supercritical regimes, the ACWs are narrowly distributed whereas in the critical regime the ACWs exhibit a broader, multimodal distribution. To quantify the diversity and disparity in intrinsic timescales across regions, we compute the entropy and range ratio of the ACW distribution. Entropy measures the dispersion of the data and the range ratio measures the ratio between the maximum and minimum values of the data. As shown in Fig.~\ref{fig:timescale}(E), both the entropy and range ratio of the ACWs increase sharply as the system approaches criticality, exhibiting a diverging trend near the critical point. These findings highlights criticality as a mechanism for the emergence of a more pronounced hierarchical structure of intrinsic timescales.

We further explore the spatial gradient of timescales across brain regions using our model by mapping the ACW to the cortical surface, as illustrated in Fig.~\ref{fig:timescale}(F). In the subcritical and supercritical regimes, ACWs across brain regions tend to be homogeneous in space. In contrast, the critical regime reveals a more distinct hierarchical organization of timescales across the cortex, with early visual areas exhibiting shorter timescales and higher-order regions showing longer timescales. In the human brain, visual processing is generally understood to occur along two distinct pathways, the ventral pathway projecting to the temporal cortex and the dorsal pathway to the parietal cortex~\cite{kravitz2013ventral,kravitz2011new}. 
Therefore, we take a closer examination at the variations in timescales along the ventral and dorsal pathways, as shown Fig.~\ref{fig:timescale}(G). 
In the subcritical and supercritical regimes, the timescales of regions along these pathways remain relatively small and similar in magnitude. However, in the critical regime, a distinctive increase in timescales is observed along both the ventral and dorsal pathways. In particular, visual sensory areas show minimal changes in timescales, while transmodal regions such as the OFC, ACC, and areas 8Ad/46 exhibit substantial increases.

\section{Discussions}

Previous works on linear network models have established the structure of whole brain anatomical connectome as a key contributor to the timescale gradient across the cortex~\cite{Chaudhuri2014eLife,Li2022pnas}. As reveal by analysis using our mean field model, this timescale gradient can be further amplified if the brain operates near the critical point of a phase transition that is absent in linear models. This result is consistent with previous findings about an optimal operating regime of the brain exhibiting neuronal cascade~\cite{Rabuffo2021eneuro,Spiegler2016eneuro,Breyton2024elife,Huang2017curr_op_neurobio}. Additional evidence is provided by comparing to timescales in a linear model embedded in human brain connectome and in our nonlinear mean field model with random connectivity (see SI). Our analysis shows that criticality and structural connectome jointly shape timescale hierarchy in the brain and more importantly indicates criticality as a unifying mechanism for diverse timescales and optimal signal transmission across the brain. These results highlight the role of dynamic interactions between brain regions in shaping timescale hierarchy and signal propagation, and suggests a potential way for the brain to self-organize toward an optimal operating regime by adjusting cross-regional coupling strength.

Previous analysis on a model of the monkey brain has attributed the emergence of timescale hierarchy to the spatial localization of eigenmodes~\cite{Li2022pnas}. However, the eigenmodes in our model do not exhibit such spatial localization and are more akin to spherical harmonics reported in~\cite{Wang2019prl,Pang2023nature}. This discrepancy could be potentially caused by the lack of directionality in DWI-derived structural connectome of the human brain, as opposed to the directed connectome of the monkey brain obtained from retrograde tracing~\cite{Markov2014cereb}.
Another contributing factor is the neuronal and synaptic nonlinearity captured by our mean field model, whose combined effect with the synaptic weight matrix determine the orthogonality of the Jacobian matrix (or the lack of it). Despite this discrepancy, our model nonetheless generates a distinct timescale gradient across the whole brain. Testing the localization of eigenmodes in the human brain requires new data or methods for estimating the directionality of structural connectome.

The human brain is the most complex physical system known to mankind. This complexity partly arises as the brain operates across a vast range of spatial and temporal scales. Although there is an abundance of models for describing the functioning of the brain at individual scales, models that can link biophysical properties of spiking neurons at the microscale to brain-wide dynamics at the macroscale are rare. 
The second-order mean field model developed in this work is derived from a spiking neural circuit model embedded with neuroimaging-derived structural connectome and it faithfully captures the nonlinear coupling between mean firing rate and firing variability. By preserving the relevant biophysical realism, our approach provides a link between microscopic and macroscopic neural dynamics missed by popular heuristic whole brain models, and offers new insights about the relationship between network structure, nonlinearity, and criticality.

\section{Methods}

\subsection{The Digital Twin Brain model}
The Digital Twin Brain (DTB) is an emerging technology aimed at simulating the human brain by incorporating known structural and functional constraints~\cite{Lu2024nsr,Lu2024nat_cs}. The DTB is implemented using a massively large spiking neural circuit model describing detailed spiking activity at single neuron resolution and is optimized through a data assimilation method. The simulation of billions of neurons and trillions of synapses in the DTB is achieved with GPU-accelerated supercomputers~\cite{Lu2024nat_cs}. Compared to conventional macroscopic models of the brain, the advantage of the DTB model lies in connecting neuronal and synaptic dynamics at fine spatiotemporal resolution (individual spikes) to experimental observations at the macroscale (whole brain imaging data).

Concretely, We consider a region-wise computational cortico-subcortical model for the whole human brain~\cite{Lu2024nsr}, comprising 378 areas based on the HCPex parcellation~\cite{huang2022extended}. Each area is represented by a random sub-network involving two spiking neural populations (excitatory and inhibitory) consisting of conductance-based leaky integrate-and-fire (LIF) neurons with synaptic decay. The ratio between excitatory and inhibitory neurons is 4:1. The relative population size of each area is derived based on regional grey matter volume (GMV). Throughout this study, we set the total number of neurons in the model to be 9.45 million.

In accordance with experimental findings, we hypothesize that the in-degree of each region is directly proportional to the sum of elements in each row of the Diffusion Weighted Imaging (DWI) matrix. We distinguish between intra-regional (local) synaptic connections and cross-regional (long range) synaptic connections, the latter of which are between excitatory populations only. We utilize a diffusion hierarchical mesoscale data assimilation (dHMDA) method~\cite{Lu2024nsr,Lu2024nat_cs} to estimate the hyperparameters of external currents ($I_{\rm ext}(t)$ in Eq.~\ref{eq:dtb_v}) to neurons in each region by fitting simulated BOLD signals to experimental BOLD signals. The simulated BOLD signals are generated using the Balloon-Windkessel model~\cite{friston2000nonlinear}. See SI for details.

\subsection{A second-order mean field approach to whole brain modeling}
\label{Moment_mapping}

While the DTB provides an invaluable computational platform for simulating the whole brain at unprecedented scales and spatiotemporal resolutions, equally important are mathematical tools for analyzing and interpreting the whole brain dynamics generated by the DTB. Our goal here is to develop a mean field model that can faithfully capture the dynamics of the DTB at macroscale.

Previous works on studying whole brain dynamics typically rely on various forms of neural mass models which describe the macroscopic interactions between nodes (brain regions)~\cite{Breakspear2017nn}. The nodes can be described by models with varying degrees of abstractions such as Kuramoto phase oscillators~\cite{Koller2024natcomms}, Stuart-Landau oscillators~\cite{Rolls2023cc}, and Wilson-Cowan-style rate models~\cite{Li2022pnas}. The main drawback of these heuristic models is that much of the microscopic properties of spiking neural dynamics are lost, namely, neuronal and synaptic nonlinearities. These nonlinearities may manifest into non-trivial effects at the whole brain level, such as phase transitions that are absent in linear models.

To preserve these biological nonlinearities in whole brain modeling, we follow a mean field approach whereby the dynamics at macroscale is derived from spiking neural dynamics on a mathematically rigorous ground. Mean field models have been widely used in studying the collective dynamics of neural populations in both local neural circuits~\cite{Brunel2000jcn,Boustani2009neco,feng2006dynamics,Zerlaut2018jcn} as well as whole brain models embedded with structural connectome~\cite{Rabuffo2021eneuro,Spiegler2016eneuro,Lavanga2023neuroimage}. Concretely, we develop a second-order mean field model based on the spiking neural circuits in the DTB using moment closure and coarse-graining. The resulting mean field model faithfully captures the nonlinear coupling between mean firing rate and firing variability to allow direct quantitative comparisons between mean field predictions and full spiking simulations of the DTB, which would be unavailable with heuristic models~\cite{Rolls2023cc,Li2022pnas}.

Let $\mathbf{m}=(\mu,\sigma^2)\in\mathbb{R}^{2n}$ be a vector representing the mean and variance of firing activity of $n$ neurons (or populations of neurons) in the sense that 
\[
\mu_i = \lim_{\Delta t\to\infty} \dfrac{\mathbb{E}[N_i(\Delta t)]}{\Delta t},\quad
\sigma_i^2 = \lim_{\Delta t\to\infty} \dfrac{{\rm Var}[N_i(\Delta t)]}{\Delta t},
\]
where $N_i$ is the spike count over a time window of $\Delta t$. Then, moment closure constitutes of the following self-consistent system of equations
\begin{equation}
    \tau\dfrac{d\mathbf{m}}{dt} = -\mathbf{m} + \Phi(\mathbf{m}),
    \label{eq:general}
\end{equation}
where $\tau$ is a time constant and $\Phi$ is a nonlinear moment mapping capturing the input-output relationship of statistical moments of an underlying spiking neural network. We refer to this model as the moment neural network (MNN). For spiking neuron models of the integrate-and-fire type, the moment closure can be derived by solving the associated first-passage time problem through a diffusion approximation~\cite{Capocelli1971Kyber,Brunel2000jcn,Boustani2009neco,feng2006dynamics}. 
Although analytical solutions to moment closure are well known, their computational complexity has prevented their applications to high-dimensional systems such as the whole brain. For the current-based LIF neuron model with instantaneous synapses, this can be overcome with an efficient numerical implementation for moment activation that we have recently developed~\cite{Qi2024pre}.

However, the activity of biological neurons is governed by conductance dynamics with diverse synaptic time scales. To capture these features, the DTB model is implemented using the conductance-based LIF neuron model with synaptic decay. The derivation of moment closure for this type of model is much more challenging due to the simultaneous presence of conductance and synaptic dynamics. To overcome this challenge, we employ an effective current approximation~\cite{Becker2024pre} which works for multiple synaptic types and the approximation error is roughly uniform across all synaptic time scales. The effective current then can seamlessly interface with the moment activation for the current-based LIF neuron model~\cite{Qi2024pre}. 
In the following, we first describe moment closure for conductance-based spiking neural circuits with synaptic decay and then coarse-graining of the DTB model.

\subsection{Moment closure for the conductance-based LIF neuron model with synaptic decay}

We begin with the conductance-based leaky integrate-and-fire neuron model with synaptic decay 
\begin{equation}
\tau_L\dfrac{dV^\alpha_i}{dt}=-(V^\alpha_i-E_L)-\sum_\beta g^{\alpha\beta} (V_i^\alpha-E_\beta)J^{\alpha\beta}_i + I_{\rm ext},
\label{eq:v}
\end{equation}
where $V^\alpha_i$ is the membrane potential of an $i$-th neuron from neural population $\alpha$, $\tau_L$ is the membrane time constant, $E_L$ is the reversal potential of the leak current, $E_\beta$ is the reversal potential for synaptic channel $\beta$, $g^{\alpha\beta}$ is the conductance, and $I_{\rm ext}$ is an external input current. The synaptic activation $J^{\alpha\beta}_i$ evoked by the pre-synaptic spike train $S_j^\beta(t)=\sum_{k}\delta(t-t^{\beta}_{j,k})$ is governed by
\begin{equation}
\dfrac{dJ^{\alpha\beta}_i}{dt}=-\dfrac{J^{\alpha\beta}_i}{\tau_\beta}+ \sum_j w^{\alpha\beta}_{ij}a^{\alpha\beta}_{ij}S_j^\beta,
\end{equation}
where $w^{\alpha\beta}_{ij}$ is the synaptic weight and $a^{\alpha\beta}_{ij}\in\{0,1\}$ indicates synaptic connectivity. When the membrane potential $V^\alpha_i(t)$ exceeds a threshold $V_{\rm th}$ it emits a spike, after which it is reset to $V_{\rm res}$ and enters a refractory period of duration $T_{\rm ref}$. For this model, the three components of the moment mappings $\Phi=\phi_{\rm MA}\circ\phi_{\rm eff}\circ\phi_{\rm sum}$ are described as follows. 

The first component $\phi_{\rm sum}$ performs a linear synaptic summation over the pre-synaptic neuronal index $j$ and it describes how moments of the pre-synaptic spike trains are transformed into that of the post-synaptic activation. Under the scenario of a large population of sparsely connected neurons, the pre-synaptic neural activity is roughly uncorrelated. Therefore, the moments of post-synaptic activation are
\begin{equation}
\tilde{\mu}^{\alpha\beta}_i = \sum_j w_{ij}^{\alpha\beta} a^{\alpha\beta}_{ij}\mu^\beta_j,
\end{equation}
\begin{equation}
(\tilde{\sigma}^{\alpha\beta}_i)^2 = \sum_j (w_{ij}^{\alpha\beta})^2 a^{\alpha\beta}_{ij}(\sigma^\beta_j)^2,
\end{equation}
where $\mu^\beta_j$ and $(\sigma^\beta_j)^2$ are the mean firing rate and firing variability of the pre-synaptic neurons, respectively.

The second component $\phi_{\rm eff}$ performs a nonlinear synaptic integration over the neural population index $\beta$ (i.e., the type of synapse) and it describes the moment mapping from the post-synaptic activation to an effective time constant and an effective current~\cite{Becker2024pre}. The effective time constant is
\begin{equation}
\bar{\tau}^\alpha_i=\dfrac{\tau_L}{1+ \sum_\beta g^{\alpha\beta} \tau_\beta \tilde{\mu}^{\alpha\beta}_i}.
\end{equation} 
The effective current mean and variance are
\begin{equation}
\bar{\mu}^{\alpha}_i=\dfrac{1}{\tau_L}(V_L+ \sum_\beta  g^{\alpha\beta}\tau_\beta E_\beta \tilde{\mu}^{\alpha\beta}_i + \mu_{\rm ext}) ,
\end{equation}
and
\begin{equation}
(\bar{\sigma}^{\alpha}_i)^2=\sum_\beta \dfrac{\bar{\tau}^\alpha_i}{\bar{\tau}^\alpha_i+\tau_\beta}h_\beta^2(\bar{E}^\alpha_i) + \dfrac{1}{\tau_L^2}\sigma_{\rm ext}^2,
\end{equation}
where $\bar{E}^\alpha_i= \bar{\tau}^\alpha_i\bar{\mu}^\alpha_i$ is the effective reversal potential, $\mu_{\rm ext}$ and $\sigma_{\rm ext}^2$ are the mean and variance of the external current $I_{\rm ext}$, and
\begin{equation}
    h_\beta(V) =\dfrac{\tau_\beta}{\tau_L}g^{\alpha\beta} (E_\beta-V)\tilde{\sigma}^{\alpha\beta}_i.
\end{equation}
Note that the current variance depends on both the moments of synaptic activation as well as the effective time constant.

The last component is the moment activation $\phi_{\rm MA}$ which is a point-wise nonlinear mapping with mean and variance given by
\begin{numcases}{}
\phi_\mu(\bar{\mu}^{\alpha}_i,\bar{\sigma}^{\alpha}_i, \bar{\tau}^\alpha_i)
= \dfrac{1}{T_{\rm ref} + 2\bar{\tau}^\alpha_i\int_{I_{\rm lb}}^{I_{\rm ub}} g(x) dx},\label{eq:ma_mu}
\\
\phi_\sigma(\bar{\mu}^{\alpha}_i,\bar{\sigma}^{\alpha}_i, \bar{\tau}^\alpha_i)
= 8(\bar{\tau}^\alpha_i)^2\mu^3\textstyle\int_{I_{\rm lb}}^{I_{\rm ub}} h(x) dx,\label{eq:ma_sigma}
\end{numcases}
Here, the functions $\phi_\mu$ and $\phi_\sigma$ are derived from the current-based LIF neuron model through a Fokker-Planck formalism and they together map the moments of the input current to that of the output spike trains~\cite{Capocelli1971Kyber,feng2006dynamics,Lu2010neuroimg}. The integration bounds are equal to 
\begin{equation}
    I_{\rm ub}(\bar{\mu}^{\alpha}_i,\bar{\sigma}^{\alpha}_i, \bar{\tau}^\alpha_i) 
= \dfrac{V_{\rm th}-\bar{\tau}^\alpha_i\bar{\mu}^{\alpha}_i}{\sqrt{\bar{\tau}^\alpha_i}\bar{\sigma}^{\alpha}_i}
\end{equation}
and
\begin{equation}
I_{\rm lb}(\bar{\mu}^{\alpha}_i,\bar{\sigma}^{\alpha}_i, \bar{\tau}^\alpha_i) 
= \dfrac{V_{\rm res}-\bar{\tau}^\alpha_i\bar{\mu}^{\alpha}_i}{\sqrt{\bar{\tau}^\alpha_i}\bar{\sigma}^{\alpha}_i}.
\end{equation}
The constant parameters $T_{\rm ref}$, $\tau_L$, $V_{\rm res}$, and $V_{\rm th}$ are identical to those in the spiking neuron model. The pair of Dawson-like functions $g(x)$ and $h(x)$ appearing in Eq.~\ref{eq:ma_mu} and Eq.~\ref{eq:ma_sigma} are $g(x)=e^{x^2}\int_{-\infty}^x e^{-u^2}du$ and $h(x)=e^{x^2}\int_{-\infty}^x e^{-u^2}[g(u)]^2du$. An efficient numerical algorithm is used to evaluate moment activation and its gradients~\cite{Qi2024pre}.

Note that the main difference from current-based LIF neuron model is that the effective time constant is a variable which depends dynamically on the synaptic inputs and is much shorter than the membrane time constant in a current-based LIF neuron model. One caveat of using the moment activation for the current-based model is the implicit assumption about the boundary condition in the stationary membrane potential distribution, i.e., $P(V_{\rm th})=0$, which is not true when synaptic decay is present. In~\cite{Becker2024pre}, it is shown that a double integration scheme can be used to estimate $P(V_{\rm th})$ by first assuming $P(E_I)=0$.

\subsection{Coarse-graining of the Digital Twin Brain model}

We apply a coarse-graining to the DTB leading to a region-wise mean field model in Eq.~\ref{eq:self-consistent equations regarding firing statistics} and Eq.~\ref{eq:Phi_decomposition}. In the following, we present results for each of the components in the moment mapping $\Phi = \phi_{\rm{MA}} \circ \phi_{\rm{eff}} \circ \phi_{\rm{sum}}$ in Eq.~\ref{eq:Phi_decomposition}.

The first component is the coarse-grained synaptic summation $\phi_{\rm sum}: \mathbb{R}^{4n}\to \mathbb{R}^{8n}$ describing the moment mapping from input spike trains to synaptic activations. For simplicity, we consider the case in which the in-degree $K^{\alpha\beta}_{uv}=\sum_j a^{\alpha\beta}_{uv,ij}$ is the same for all neurons in a given region. The mean and variance of the synaptic activation can then be written as
\begin{equation}
\tilde{\mu}^{\alpha\beta}_u =  w_{uu}^{\alpha\beta}K_{uu}^{\alpha\beta}\mu_u^{\beta} 
+\delta_{\alpha,E}\delta_{\beta,E}\sum_{v\neq u} w_{uv}^{EE}K_{uv}^{EE}\mu^E_v,
\end{equation}
and
\begin{equation}
(\tilde{\sigma}^{\alpha\beta}_u)^2 =  (w_{uu}^{\alpha\beta})^2K_{uu}^{\alpha\beta}(\sigma^{\beta}_u)^2 
+ \delta_{\alpha,E}\delta_{\beta,E}\sum_{v\neq u} (w_{uv}^{EE})^2K_{uv}^{EE}(\sigma^{E}_v)^2,
\end{equation}
respectively. The first term in each of these equations accounts for the intra-regional synaptic coupling whereas the second term accounts for the cross-regional synaptic coupling which applies only to excitatory neurons.

The second component is the effective current $\phi_{\rm eff}:\mathbb{R}^{8n}\to\mathbb{R}^{6n}$. To get this, we first calculate the effective time constant
\begin{equation}
    \bar{\tau}_u^\alpha=\dfrac{\tau_{L,\alpha}}
    {1+\sum_\beta g_{u}^{\alpha\beta}\tau_\beta\tilde{\mu}^{\alpha\beta}_{u}}.
\end{equation}
The mean and variance of the effective current can then be calculated as
\begin{equation}
    \bar{\mu}_u^\alpha = \dfrac{1}{\tau_{L,\alpha}}(
    E_L + \sum_\beta g_{u}^{\alpha\beta}\tau_\beta E_\beta\tilde{\mu}^{\alpha\beta}_{u} +\mu_{\rm bg}^\alpha + \mu^\alpha_{u,\rm ext}
    ),
    \label{eq:rw_mnn_eff_curr_mean}
\end{equation}
and
\begin{equation}
    (\bar{\sigma}_u^\alpha)^2 =
\sum_\beta \dfrac{\bar{\tau}^\alpha_u}{\bar{\tau}^\alpha_u+\tau_\beta}(h^\alpha_{u,\beta})^2 
+\dfrac{\bar{\tau}^\alpha_u}{\bar{\tau}^\alpha_u+\tau^\alpha_{\rm bg}}(h^\alpha_{\rm bg})^2 
    +\dfrac{1}{\tau_{L,\alpha}^2}(\sigma^\alpha_{u,\rm ext})^2,
    \label{eq:rw_mnn_eff_curr_var}
\end{equation}
respectively. Let $\bar{E}^\alpha_u= \bar{\tau}^\alpha_u\bar{\mu}^\alpha_u$ be the effective reversal potential. We also have
\begin{equation}
    h^\alpha_{u,\beta} =\dfrac{\tau_\beta}{\tau_{L,\alpha}}g_{u}^{\alpha\beta} (E_\beta-\bar{E}^\alpha_u)\tilde{\sigma}^{\alpha\beta}_u,
\end{equation}
and $h^\alpha_{\rm bg}=\sqrt{2\tau^\alpha_{\rm bg}}\sigma^\alpha_{\rm bg}/\tau_{L,\alpha}$. Here $\mu^\alpha_{\rm bg}$ and $(\sigma^\alpha_{\rm bg})^2$ are the mean and variance of a local background current modeled as an OU process, whereas $\mu^\alpha_{u,\rm ext}$ and $(\sigma^\alpha_{u,\rm ext})^2$ are the mean and variance of an external current, interpreted as being originated from sensory inputs. The external input mean $\mu_{u,\rm ext}^\alpha$ in Eq.~\ref{eq:rw_mnn_eff_curr_mean}-\ref{eq:rw_mnn_eff_curr_var} is set according to the data-assimilated current whereas the external input variance is set to $\sigma_{u,\rm ext}^\alpha=0$. Specifically, for results presented in Fig.~\ref{fig:fig2} and Fig.~\ref{fig:bold_compare_dtb_MNN}, $\mu_{u,\rm ext}^\alpha$ varies slowly over time and are constants within 800 ms time windows; for all remaining results, the average value over all time windows is used.

For the third component $\phi_{\rm MA}: \mathbb{R}^{6n}\to\mathbb{R}^{4n}$, we apply the moment activation to the effective current received by each brain region to get
\begin{equation}
    \phi_\mu = \phi_\mu(\bar{\mu}^\alpha_u, \bar{\sigma}^\alpha_u,\bar{\tau}^\alpha_u),
\end{equation}
\begin{equation}
    \phi_\sigma = \phi_\sigma(\bar{\mu}^\alpha_u, \bar{\sigma}^\alpha_u,\bar{\tau}^\alpha_u).
\end{equation}
Since the moment activation is a point-wise operation, the moment mappings $\phi_\mu$ and $\phi_\sigma$ in the coarse-grained model are identical to the fine-grained version (Eq.~\ref{eq:ma_mu}-\ref{eq:ma_sigma}) but with the neuronal index $i$ replaced with regional index $u$. This is valid under the scenario that the neural circuit within each brain region is homogeneous.

The time constant $\tau$ in the mean field model is calibrated by matching the oscillation frequency in the mean field model to that of the population firing rate in the DTB model, just above their respective critical points. We set the synaptic weights of intra-regional connections to $w^{\alpha\beta}_{uu}=1$ and those of cross-regional connections to $w^{{\rm EE}}_{uv}=\gamma$ when $u\neq v$, where $\gamma$ is a global factor controlling the cross-regional synaptic coupling strength. The values of all remaining parameters of the DTB and the mean field model are specified in Table~\ref{tab:default parameters}.

\begin{table}
\centering
\renewcommand{\arraystretch}{1.2} 
\begin{tabular}{clclll}
\multicolumn{2}{c}{\cellcolor[HTML]{AEAAAA}Neuronal} & \multicolumn{2}{c}{\cellcolor[HTML]{AEAAAA}Synaptic} & \multicolumn{2}{c}{\cellcolor[HTML]{AEAAAA}Background input} \\
 \multicolumn{1}{c}{$E_{\rm L}$}            & \multicolumn{1}{l}{-70 mV}  &      \multicolumn{1}{c}{$E_{\rm E}$}                 & \multicolumn{1}{l}{0 mV}    &          \multicolumn{1}{c}{$\mu_{\rm bg}^{\rm E}/\mu_{\rm bg}^{\rm I}$}             & \multicolumn{1}{l}{7.2/5.4 mV}   \\
      \multicolumn{1}{c}{$V_{\rm res}$}                   & \multicolumn{1}{l}{-60 mV}   &   \multicolumn{1}{c}{$E_{\rm I}$}                     & \multicolumn{1}{l}{-70 mV}  & \multicolumn{1}{c}{$\sigma_{\rm bg}^{\rm E}/\sigma_{\rm bg}^I$}    & \multicolumn{1}{l}{20/12 mV}  \\
\multicolumn{1}{c}{$V_{\rm th}$}  &                 \multicolumn{1}{l}{-50 mV}             & \multicolumn{1}{c}{$\tau_{\rm E}$}  &                  \multicolumn{1}{l}{4 ms}           &                  \multicolumn{1}{c}{$\tau_{\rm bg}$}         &         \multicolumn{1}{l}{4 ms}                      \\
\multicolumn{1}{c}{$T_{\rm ref}^{\rm E}/T_{\rm ref}^I$}  &            \multicolumn{1}{l}{2/1 ms}                 &  \multicolumn{1}{c}{$\tau_{\rm I}$} &         \multicolumn{1}{l}{10 ms}                     &                  &     \\
\multicolumn{1}{c}{$\tau_{\rm L,\rm E}/\tau_{\rm L,\rm I}$}  &            \multicolumn{1}{l}{20/10 ms}                 &  \multicolumn{1}{l}{} &         \multicolumn{1}{c}{}                     &                  & \\

\multicolumn{1}{c}{$\hat{g}^{EE}/\hat{g}^{IE}$}  &            \multicolumn{1}{l}{0.0025/0.0275}                 &  \multicolumn{1}{l}{} &         \multicolumn{1}{c}{}                     &                  & \\
\multicolumn{1}{c}{$\hat{g}^{EI}/\hat{g}^{II}$}  &            \multicolumn{1}{l}{0.108/0.024}                 &  \multicolumn{1}{l}{} &         \multicolumn{1}{c}{}                     &                  & 
\end{tabular}
\caption{\label{tab:default parameters} Default parameter values for the DTB and the mean field models. Parameters without superscripts indicate that they are the same for both excitatory and inhibitory populations, while parameters with superscripts E or I indicate different values for the excitatory and inhibitory populations.}
\end{table}

\subsection{Linear stability analysis}
The external input currents are set to constants according to the time averages of the assimilated $I_{\rm ext}$ during resting state. Both cortical and subcortical regions are included for this analysis. Fixed points $\mathbf{m}_0$ of the system defined by Eq.~\eqref{eq:self-consistent equations regarding firing statistics} are found using the Newton-Raphson method and a linearization is applied to the system around the fixed points $\mathbf{m}=\mathbf{m}_0+\epsilon \tilde{\mathbf{m}}$ to obtain
\begin{equation}
    \dfrac{d\tilde{\mathbf{m}}}{dt} = \boldsymbol{J}\tilde{\mathbf{m}},
\end{equation}
where $\boldsymbol{J}$ is the Jacobian matrix of the system 
\begin{equation}
    \boldsymbol{J}= \frac{1}{\tau} \left(-\mathbf{I}_{4n}+\boldsymbol{J}_{\rm{MA}}\boldsymbol{J}_{\rm{eff}}\boldsymbol{J}_W \right).
\end{equation}
Here, $\boldsymbol{J}_{\rm{MA}}$, $\boldsymbol{J}_{\rm{eff}}$, and $\boldsymbol{J}_W$ are the gradients of the moment mappings $\phi_{\rm MA}$, $\phi_{\rm eff}$ and $\phi_{\rm sum}$ respectively. In practical implementation, we use PyTorch's automatic differentiation to obtain the gradients $\boldsymbol{J}_W$ and $\boldsymbol{J}_{\rm{eff}}$ of the mappings $\boldsymbol{\phi}_{\rm{sum}}$ and $\boldsymbol{\phi}_{\rm{eff}}$, respectively. Additionally, we analytically derive the gradient $\boldsymbol{J}_{\rm{MA}}$ of the moment activation $\boldsymbol{\phi}_{\rm{MA}}$ (see SI). 
We numerically solve for the eigenvalues and eigenmodes of the Jacobian matrix and sort the eigenvalues in descending order based on their real parts. The stability of the system is then indicated by the sign of the real parts of the eigenvalues, whereas the angular frequency of the oscillation is indicated by the imaginary parts. 

\subsection{Quantifying signal attenuation of stimulus-evoked response}

Here we outline procedure of the numerical experiments, as illustrated in Fig.~\ref{fig:Enhancement_propagation_power}(A). Initially, we simulate the model over an extended period of time for it to reach a steady state, characterized by a stable fixed point in the subcritical regime and a limit cycle in the supercritical regime. Following this, we apply a 100 ms, 20 mV current stimulus to V1 in both hemispheres and analyze the response signals of different brain regions. The same current stimulus is applied as cross-regional coupling strength $\gamma$ is varied. Here we focus on the relaxation of mean firing rate of excitatory populations in different brain regions after stimulus removal. 
In the supercritical state, the input stimulus induces a phase shift $T_{\rm lag}$ to the limit cycle. To numerically estimate the phase shift, we minimize a loss function $\max_{t} |\mu_{\rm E}^{\rm post}(t)-\mu_{\rm E}^{\rm pre}(t-T_{\rm lag})|$ over $T_{\rm lag}$. We simulate the model for at least 0.5 seconds after stimulus removal to guarantee convergence to the steady-state limit cycle. System's response is then calculated as the difference between the phase-adjusted response trajectory and the unperturbed trajectory. The model is simulated using a first-order Euler method with a time step of 0.001 ms.

\subsection{Quantifying intrinsic neural timescales}

The way intrinsic neural timescales are estimated varies significantly across the literature. The specific method used depends on whether the recorded neural activity is evoked or on-going. For evoked activity, intrinsic timescales are measured from the rise and decay time of neural response to transient stimulus~\cite{Rolls2023cc}. For on-going activity, intrinsic timescales are measured from the decay time of autocorrelation function (ACF) of neural activity fluctuations~\cite{Manea2022elife,Watanabe2019eLife,Ito2020neuroimage,Murray2014nn,Cavanagh2016elife,Zeraati2023natcomms,Rossi-Pool2021npas,Golesorkhi2021b,Watanabe2019eLife,Manea2022elife} or from the power spectral density (PSD)~\cite{Donoghue2020nn,Gao2020eLife}. We provide a critical assessment of different ways intrinsic neural timescales are estimated in experimental and modeling studies in SI and draw conceptual links between them using impulse response theory.

In this work, we consider the temporal correlation of neural response to persistent input stimulus to V1, modeled as Gaussian white noise. We require a consistent method for quantifying timescales for both subcritical and supercritical regimes. Unfortunately, ACF of the raw response signal is ill-suited for the supercritical regime. To overcome this, we adopt an envelope method frequently used for quantifying temporal correlations in experimental recordings of oscillatory neural activity~\cite{Linkenkaer-Hansen2001JoNeur,Palva2013pnas}. Specifically, for the subcritical and critical dynamical regimes, we compute the envelope (magnitude of the analytic signal) of the neural response in each region. For the supercritical regime, we first estimate the oscillatory frequency (using a representative brain region, 8Ad in the right hemisphere) and then apply band-pass filtering with a rectangular window of width 10 Hz centered around this frequency to the neural response in each region, before calculating the envelope. We quantify the intrinsic timescales using the half-life of the ACF of the envelope.

\section*{Acknowledgments}

This work is supported by National Science and Technology Major Project (No. 2018AAA0100303); partially supported by the Science and Technology Commission of Shanghai Municipality (No. 23ZR1403000) and the National Natural Science Foundation of China (No. 12471481; No. 62306078). This work has benefited from funding of the European Union’s Horizon Europe Programme under the Specific Grant Agreement No. 101137289 (Virtual Brain Twin Project) and No. 101147319 (EBRAINS 2.0). We thank Songting Li for the useful discussions and suggestions.

\section*{Code and data availability}
The source data for the Digital Twin Brain project is available at: \url{https://doi.org/10.6084/m9.figshare.27310665}. The code for the second-order mean field model is available at: \url{https://github.com/BrainsoupFactory/moment-neural-network}.

\section*{Author Contributions}
Conceptualization: Y.Q. and J.F.; Methodology: Y.Q., W.D., W.L. and J.F.; Investigation: Y.Q., J.W.; Software: Y.Q., J.W.; Visualization: Y.Q. and J.W.; Writing — original draft: Y.Q. and J.W.; Writing — review \& editing: Y.Q., J.W., W.D., D.G., V.J., W.L., and J.F.; Supervision: J.F.


\bibliographystyle{unsrt}

\newpage

\begin{flushleft}
Supplementary Information\\
{\Large
\textbf
\newline{Optimal signal transmission and timescale diversity in a model of human brain operating near criticality}
}
\newline
\end{flushleft}

\setcounter{section}{0}
\renewcommand{\thesection}{S\arabic{section}}
\setcounter{equation}{0}
\renewcommand{\theequation}{S\arabic{equation}}
\setcounter{figure}{0}
\renewcommand{\thefigure}{S\arabic{figure}}
\setcounter{table}{0}
\renewcommand{\thetable}{S\arabic{table}}

\setcounter{page}{1}

\section{Construction of the DTB model}
Here, we outline the construction of the DTB model used in this work and refer to our previous paper~\cite{Lu2024nsr,Lu2024nat_cs} for additional backgrounds. 

\subsection{A spiking cortico-subcortical model of whole human brain}\label{sec:Networkmodel}

We describe details of the construction of the region-wise computational cortico-subcortical model~\cite{Lu2024nsr,Lu2024nat_cs}. The model comprises $n=378$ areas based on the HCPex parcellation~\cite{huang2022extended}. Each area is represented by a random sub-network involving excitatory and inhibitory populations. The ratio between the number of excitatory and inhibitory neurons is 4:1. Concretely, denote $i,j\in\{0,1,N_u-1\}$ the neuronal index within each region, $\alpha,\beta\in \{{\rm E},{\rm I}\}$ the index for the neuronal type, and $u,v\in \{0,1,\dots,n-1\}$ the brain region index, we can write down this model as
\begin{equation}
\tau_{L,\alpha}\dfrac{dV^\alpha_{u,i}}{dt}=-(V^\alpha_{u,i}-E_L)-\sum_{\beta} g_{u}^{\alpha\beta} (V_{u,i}^\alpha-E_\beta)J^{\alpha\beta}_{u,i} + I^\alpha_{u,i,\rm bg} + I^\alpha_{u,i,\rm ext},
\label{eq:dtb_v}
\end{equation}
\begin{equation}
\dfrac{dJ^{\alpha\beta}_{u,i}}{dt}=-\dfrac{J^{\alpha\beta}_{u,i}}{\tau_\beta}
+ \sum_j w^{\alpha\beta}_{uu,ij}a^{\alpha\beta}_{uu,ij}S_{u,j}^\beta
+ \delta_{\alpha,{\rm E}}\delta_{\beta,{\rm E}}\sum_{v\neq u,j} w^{{\rm EE}}_{uv,ij}a^{{\rm EE}}_{uv,ij}S_{v,j}^{\rm E},
\label{eq:dtb_J}
\end{equation}
\begin{equation}
\tau^\alpha_{\rm bg} dI^\alpha_{u,i,{\rm bg}} = (\mu^\alpha_{\rm bg} - I^\alpha_{u,i,{\rm bg}})dt + \sqrt{2\tau^\alpha_{\rm bg}} \sigma^\alpha_{\rm bg} dW_{t},
\label{eq:dtb_curr}
\end{equation}
which is a special case of Eq.~\ref{eq:v}. The synaptic connectivity $a^{{\rm EE}}_{uv,ij}$ within and between regions are drawn randomly with constant in-degrees $K_{uv}^{\alpha\beta}$ only depending on the identities of source and target populations. 
In addition, a background input current $I^\alpha_{u,i,\rm bg}$ is modeled as an OU process whose parameters $\mu^\alpha_{\rm bg}$, $\sigma^\alpha_{\rm bg}$ and $\tau^\alpha_{\rm bg}$ are constants. The external inputs $I^\alpha_{u,i,\rm ext}(t)$ are determined via a data assimilation approach and they vary slowly over time (constants over 800 ms time windows).

To set the number of neurons, we first specify the total number of neurons $N=9,450,000$ of the entire brain and then set the number of neurons in each region $N_u$ in proportion to the regional gray matter volume. The number of neurons is cropped such that $N_u>2500$ in all regions. To set the in-degrees, we first specify the value of a parameter $K=500$ controlling the average in-degree of cross-regional connections. The total number of cross-regional connections is then $KN$ and the total number of cross-regional connections from a source region to a target region is set in proportion to the diffusion-weighed imaging (DWI) tractography. This is then divided by the number of neurons in the target region $N_u$ to get the (excitatory) cross-regional in-degrees $K_{uv}$. We crop the cross-regional in-degree $K_u=\sum_v K_{uv}$ of each neuron in a target region to the range $[71,3500]$. This value is then used to estimate the population-specific in-degrees according to the following scaling. For intra-regional (local) connections, we have 
$K_{uu}^{\alpha {\rm E}}\leftarrow 0.6K_u$ and 
$K_{uu}^{\alpha {\rm I}}\leftarrow 0.15K_u$; for cross-regional (long range) connections, we have $K_{uv}^{\rm EE}\leftarrow 0.25K_u$. The final values of the population-specific in-degree values are provided as Supplemental Data. To keep neuronal firing to be within biological range, the conductance $g_{u,\beta}$ of the $u$-th region is calculated as
\begin{equation}
    g_{u}^{\alpha\beta} = \hat{g}^{\alpha\beta} \tfrac{K}{K_u},\quad \beta\in\{\text{E,I}\},
\end{equation}
where $\hat{g}^{\alpha\beta}$ is a unit conductance. External currents $I^{\rm E}_{u,i,\rm ext}(t)$ to excitatory populations model sensory input from the environment and follows a Gamma distribution with a shape parameter $\alpha=5$ and a rate parameter $\lambda_{u,\rm ext}(t)$, the latter of which is determined from the experimental BOLD signal through data assimilation.

For simplicity, we set the synaptic weights of intra-regional connections to $w^{\alpha\beta}_{uu,ij}=1$ and those of cross-regional connections to $w^{{\rm EE}}_{uv,ij}=\gamma$ when $u\neq v$, where $\gamma$ is a global factor controlling the cross-regional synaptic coupling strength. The values of model parameters are displayed in Table \ref{tab:default parameters}.

\subsection{BOLD model}

The BOLD model has two components, regional cerebral blood flow (rCBF) and blood-oxygenation-level-dependent (BOLD) signal. First, the BOLD component. The BOLD signal $y(t)$ given the blood inflow $f_{\rm in}(t)$ follows
\begin{align}
y(t)&= V_0[k_1(1-q)+k_2(1-q/v)+k_3(1-v)],\\
\tau_0 \dfrac{dv}{dt}&= f_{\rm in} - f_{\rm out}(v),\\
\tau_0 \dfrac{dq}{dt}& = Q(f_{\rm in}) - f_{\rm out}(v)q/v,
\end{align}
where $Q(f_{\rm in})=\tfrac{f_{\rm in}}{E_0}[1-(1-E_0)^{1/f_{\rm in}}]$, $f_{\rm out}(v)=v^{1/\alpha}$, $\alpha=0.2$, $k_1=7E_0$, $k_2=2$, $k_3=2E_0-0.2$, $V_0=0.02$, $\tau_0=1$ and $E_0=0.8$. The input is the blood inflow $f_{\rm in}$. 
Second, the rCBF component. The blood inflow $f_{\rm in}(t)$ given neural activity $r(t)$ follows
\begin{align}
\dfrac{df_{\rm in}}{dt}&=s,\\
\dfrac{ds}{dt}&= -s/\tau_s - (f_{\rm in}-1)/\tau_f +\epsilon r(t),
\label{eq:rCBF}
\end{align}
where $\tau_s=0.8$, $\tau_f=0.4$ and $\epsilon=1$.
 
Simulated BOLD signals in the mean field model are calculated as follows. We first randomly sample the instantaneous firing rate $x_{u,i}(t)$ for each neuron $i$ in brain region $u$ from the Gaussian distribution specified by the mean field model. Then, the population firing rate for brain region $u$ can be calculated as
\[
r_u(t) = \dfrac{\sum_i x_{u,i}(t)}{N_u},
\]
which is then fed into Eq.~\ref{eq:rCBF} to calculate the BOLD signal for each brain region.

\subsection{Diffusion hierarchical mesoscale data assimilation (dHMDA)} \label{sec:dHMDA}
In this study, we utilize the diffusion hierarchical mesoscale data assimilation (dHMDA)~\cite{Lu2024nat_cs} to estimate the hyperparameters of external currents for neurons associated with each region. This estimation process involves fitting the simulated BOLD signals to experimental BOLD signals, with the simulated BOLD signals generated using the Balloon-Windkessel model~\cite{friston2000nonlinear}.

In our approach, each region representing a two-population sub-network is designated as a region of interest (ROI). We assume that excitatory neurons of the same type within each ROI share a common set of parameters $\lambda_{u,\rm ext}(t)$. We utilize the diffusion ensemble Kalman filter for parameter inference, leveraging its capacity to effectively manage high-dimensional observations and mitigate the inherent challenges posed by the limited number of data time points. The hyperparameter inference is executed through a straightforward random-walk bootstrap filter~\cite{Lu2024nat_cs}. Analysis and prediction using the dHMDA filter is executed at each time point of the BOLD signals, with the time scale synchronized with the biological clock (in milliseconds). Each time step corresponds to the period of fMRI scanning. Following the estimation of hyperparameters for a sub-network using an observation signal (experimental BOLD signal), we re-simulate this model by assigning values of the external current to each neuron based on the hyperparameter times series and assess the goodness of fit.

In practical implementation, the hyperparameters of the external currents to neurons of the ROI are estimated by the dHMDA in the cortico-subcortical model of the DTB with 9 million neurons and an average in-degree of 500. Subsequently, we collect population-level spike rates from each ROI and compute the region-level hemodynamics to derive the BOLD signals. We conduct iterations of the dHMDA process to adjust the hyperparameters and then proceed to resample and update neural parameters based on the hyperparameters. In this way, the assimilation process successfully aligns the simulated BOLD signal of the model with experimental data.

\section{Gradient of the moment activation}

For the linear stability analysis, it is necessary to calculate the gradient of the moment activation for obtaining the Jacobian matrix. The gradient of moment activation for fixed membrane time constant has been presented previously~\cite{Qi2024pre}. For conductance-based LIF neuron with synaptic decay, the effective current approximation leads to state-dependent time constants which must be taken into account in calculating the Jacobian matrix. The partial derivatives of the moment activation are calculated from Eq.~\ref{eq:ma_mu} and Eq.~\ref{eq:ma_sigma} as
\begin{equation}
    \begin{aligned}
\displaystyle \frac{\partial \mu}{\partial \bar{\mu}} =& 2 \bar{\tau}\sqrt{\bar{\tau}} \frac{\mu^2}{\bar{\sigma}} [g(I_{\text{ub}}) - g(I_{\text{lb}})],\\
\displaystyle \frac{\partial \mu}{\partial \bar{\sigma}} =& 2 \bar{\tau} \frac{\mu^2}{\bar{\sigma}} [g(I_{\text{ub}})I_{\text{ub}} - g(I_{\text{lb}})I_{\text{lb}}],\\
\displaystyle \frac{\partial \mu}{\partial \bar{\tau}} 
=& -\mu^2 \left\{2 [G(I_{\text{ub}}) - G(I_{\text{lb}})] +  \left[g(I_{\text{ub}})\frac{-V_{\text{th}}-\bar{\tau}\bar{\mu}}{\bar{\sigma} \sqrt{\bar{\tau}}} - g(I_{\text{lb}})\frac{-V_{\text{res}}-\bar{\tau}\bar{\mu}}{\bar{\sigma} \sqrt{\bar{\tau}}}\right]
\right\}.
    \end{aligned}
\end{equation}
and 
\begin{equation}
    \begin{aligned}
\displaystyle \frac{\partial \sigma}{\partial \bar{\mu}} =& 3\bar{\tau}\sqrt{\bar{\tau}} \frac{\sigma}{\bar{\sigma}}\mu [g(I_{\text{ub}}) - g(I_{\text{lb}})]-\frac{1}{2}\sqrt{\bar{\tau}}\frac{\sigma}{\bar{\sigma}} \frac{h(I_{\text{ub}}) - h(I_{\text{lb}})}{H(I_{\text{ub}}) - H(I_{\text{lb}})},\\
\displaystyle \frac{\partial \sigma}{\partial \bar{\sigma}} = &3\bar{\tau} \frac{\sigma}{\bar{\sigma}}\mu [g(I_{\text{ub}})I_{\text{ub}} - g(I_{\text{lb}})I_{\text{lb}}]-\frac{1}{2}\frac{\sigma}{\bar{\sigma}} \frac{h(I_{\text{ub}})I_{\text{ub}} - h(I_{\text{lb}})I_{\text{lb}}}{H(I_{\text{ub}}) - H(I_{\text{lb}})}& ,\\
\displaystyle \frac{\partial \sigma}{\partial \bar{\tau}} =& \frac{3}{2} \frac{\partial \mu}{\partial \bar{\tau}} \frac{\sigma}{\mu} +  \frac{\sigma}{\bar{\tau}} + \frac{\sigma}{2} \frac{1}{H(I_{\text{ub}}) - H(I_{\text{lb}})} \left[h(I_{\text{ub}})\frac{-V_{\text{th}}-\bar{\tau}\bar{\mu}}{2\bar{\sigma} \bar{\tau}\sqrt{\bar{\tau}}} -h(I_{\text{lb}})\frac{-V_{\text{res}}-\bar{\tau}\bar{\mu}}{2\bar{\sigma} \bar{\tau}\sqrt{\bar{\tau}}}\right].
    \end{aligned}
\end{equation}
The derivatives have removable singularities as $\bar{\sigma}\to 0$. In this scenario, we calculate the limits of the partial derivative of $\phi_\mu$ and $\phi_\sigma$ as
\begin{equation}
    \begin{aligned}
\displaystyle &\lim_{\bar{\sigma} \to 0} \frac{\partial \mu}{\partial \bar{\mu}} = \left\{
\begin{array}{rl}
\displaystyle 0, & \text{for } \bar{\mu}\bar{\tau} \le V_{\text{th}}\\
\displaystyle \frac{V_{\text{th}}\bar{\tau}\mu^2}{\bar{\mu}(\bar{\mu}\bar{\tau}-V_{\text{th}})}, & \text{for } \bar{\mu}\bar{\tau} > V_{\text{th}}.
\end{array} \right. \\
\displaystyle &\lim_{\bar{\sigma} \to 0}\frac{\partial \mu}{\partial \bar{\sigma}} =0 ,\\
\displaystyle &\lim_{\bar{\sigma} \to 0} \frac{\partial \mu}{\partial \bar{\tau}} 
= \left\{
\begin{array}{rl}
\displaystyle 0, & \text{for } \bar{\mu}\bar{\tau} \le V_{\text{th}}\\
\displaystyle \mu^2 \left[\log \left(1-\frac{V_{\text{th}}}{\bar{\mu}\bar{\tau}}\right) 
 + \frac{V_{\text{th}}}{\bar{\mu}\bar{\tau}-V_{\text{th}}} \right], & \text{for } \bar{\mu}\bar{\tau} > V_{\text{th}}.
\end{array} \right.
    \end{aligned}
\end{equation}
and
\begin{equation}
    \begin{aligned}
\displaystyle \lim_{\bar{\sigma} \to 0} \frac{\partial \sigma}{\partial \bar{\mu}} =&  0,\\
\displaystyle \lim_{\bar{\sigma} \to 0} \frac{\partial \sigma}{\partial \bar{\sigma}} =&  \sqrt{\frac{\bar{\tau}}{2}} \mu^{\frac{3}{2}}\sqrt{\frac{\bar{\tau}^2}{(V_{\text{th}}-\bar{\mu}\bar{\tau})^2} - \frac{1}{\bar{\mu}^2}},\\
\displaystyle \lim_{\bar{\sigma} \to 0} \frac{\partial \sigma}{\partial \bar{\tau}} =&  0.
\end{aligned}
\end{equation}

\section{Quantifying intrinsic neural timescales}\label{sec:Quantifying intrinsic neural timescales}

Given the inconsistency in the definition of intrinsic timescales across the literature, we provide a critical assessment of how they are estimated or defined in experimental and modeling studies. We then formulate intrinsic timescales from the perspective of impulse response theory which offers a conceptual link between different ways of measuring intrinsic timescales. Finally, we point out the limitations of existing approaches and propose an improved way of measuring intrinsic timescales based on signal envelope.

The way intrinsic timescales is measured in biological experiments depends on whether the recorded neural activity is evoked or on-going. In the first scenario, a transient stimulus is presented to the animal or human participant and is then removed after a brief period (typically around a hundred milliseconds). Intrinsic timescales can then be measured from the rise and decay of the stimulus-evoked activity~\cite{Rolls2023cc}. On-going cortical activity may be recorded when the brain is in resting state~\cite{Manea2022elife,Watanabe2019eLife,Ito2020neuroimage}, in an attentive state before stimulus onset~\cite{Murray2014nn,Cavanagh2016elife,Zeraati2023natcomms,Rossi-Pool2021npas}, or during the delay period after stimulus presentation but before taking an action~\cite{Gao2020eLife}. 
In all cases, the autocorrelation function (ACF) may be calculated from on-going neural activity, with an implicit assumption that the activity is roughly stationary. The intrinsic timescales can then be estimated from the ACF by fitting it to an exponential function (or multiple exponentials), or by its time to reaching half maximum. In some cases, the ACF may overshoot to negative values before decaying towards zero and the 0-crossing time is used as a measure of intrinsic timescales~\cite{Golesorkhi2021b}. A number of studies have used the area-under-the-curve before 0-crossing as an index for intrinsic timescales~\cite{Watanabe2019eLife,Manea2022elife}. 
Sometimes, on-going neural activity is described using its power spectral density (PSD) and it is possbile to estimate the intrinsic timescales from the `knee' frequency of the aperiodic component of the PSD~\cite{Donoghue2020nn,Gao2020eLife}. In studies of long range temporal correlations, the envelope of on-going oscillatory neural activity is first calculated~\cite{Linkenkaer-Hansen2001JoNeur,Palva2013pnas}, whose autocorrelation is used to quantify (potentially diverging) timescales.

Here, we briefly discuss the conceptual link between different ways of measuring intrinsic timescales from the perspective of linear response theory. Consider a general nonlinear dynamical system defined by
\[
\tau\dfrac{d\mathbf{m}}{dt} = -\mathbf{m} + \Phi(\mathbf{m}; \theta)
\]
where $\mathbf{m}$ represents the neural activity state and $\theta$ is an arbitrary parameter of interest, e.g., an input current to V1. Here, we are interested in the response of the system to an impulse perturbation (delta function) to the parameter $\theta$ near its stable fixed point. The impulse response function of the linearized system is
\begin{equation}
h(t)= \mathcal{L}^{-1}\left\{ [(1+\tau s)I - \nabla_{\mathbf{m}}\Phi]^{-1} \partial_\theta \Phi \right\},
\label{eq:imp_resp_soln}
\end{equation}
where $\mathcal{L}^{-1}$ denotes the inverse Laplace transform and $\nabla_{\mathbf{m}}\Phi$ the Jacobian matrix of the system evaluated at the fixed point. The impulse response function (i.e., Green's function) describes the system's dynamics near the fixed point under impulse perturbations to its parameters. 

The impulse response function can then be used to connect different ways of measuring intrinsic timescales. If we model sensory input to a neural system as an arbitrary temporal perturbation $\delta \theta$, then stimulus-evoked activity can be calculated as a convolution with the impulse response 
$\delta\mathbf{m}(t)=[h\ast \delta \theta](t) = \int_0^\infty h(t-t')\delta\theta(t')dt'$. 
Impulse response theory also provides a conceptual link between the stimulus-evoked response and ACF. In general, if an input $X(t)$ with ACF $R_X(\tau)$ passes through an linear time-invariant system to yield output $Y(t)$, then the ACF of $Y(t)$ is $R_Y(\tau) = \int_{-\infty}^\infty\int_{-\infty}^\infty h(s)h(r)R_X(\tau+s-r)dsdr$, where $h(t)$ is the two-sided impulse response function. In the special case where the input $X(t)$ is a Gaussian white noise the input ACF is $R_X(\tau)=\delta(\tau)$, and the corresponding ACF of the output simplifies to
$R_Y(\tau)=\int_{-\infty}^\infty h(s)h(\tau+s)ds$. 
The ACF can be linked to the power spectral density (PSD) of a signal as a Fourier transform pair according to the Wiener–Khinchin theorem $S(\omega) = \int_{-\infty}^\infty e^{-i\omega\tau} R_Y(\tau)d\tau$, assuming the signal is stationary. This relationship has been previously used to estimate intrinsic timescales from the PSD of on-going neural activity~\cite{Gao2020eLife}.

There are a couple of issues with how intrinsic timescales is quantified in previous studies. Firstly, in a number of studies, the ACF exhibits overshoot to negative values before decaying to zero. This could be a sign of oscillations in the underlying neural activity that is blurred away by temporal filtering (e.g. hemodynamic response in fMRI recording) or trial averaging. In this scenario, measurement based on half-life or time at 0-crossing cannot distinguish effects from oscillations and decay. Secondly, quantifying intrinsic timescales from ACF discards information about the strength of signal propagated to different parts of a system. Here, we argue that to properly quantify intrinsic timescales, we should jointly account for oscillations, decay timescales, and energy of signal transmission.

To this end, we propose that a better way of defining intrinsic timescales is through the envelope of a signal. If $s(t)$ is a signal (either stimulus-evoked response or on-going activity), then its analytic signal is defined as $a(t)=s(t)+i\hat{s}(t)$, where $\hat{s}(t)$ denotes the Hilbert transform of $s(t)$. We can then calculate the envelope of the signal as $\lvert s(t)\rvert$. The autocorrelation function of the envelope fluctuations of on-going activity in each brain region can then be calculated, from which the timescales can be estimated by half-life or fitting to exponential functions. 
The advantage of this method is its applicability to both the subcritical regime with damped oscillations and the supercritical regime with limit cycles. 

 \section{Nonlinearity and structural connectome jointly shape timescale hierarchy}

As demonstrated in the main text, when operating in the critical regime, our mean field model displays more diverse and distinct temporal dynamics during visual signal processing and reveals a clearer hierarchical organization of intrinsic timescales along the visual processing pathways. Given that the model incorporates two essential features of the brain, biophysical nonlinearity and structural connectome, we seek to further dissect the contributions from each of these components to the diversity and disparity of intrinsic timescales in visual signal processing. For this purpose, we conduct two sets of control experiments as follows.

\begin{figure}
    \centering
\includegraphics[width=1\textwidth]{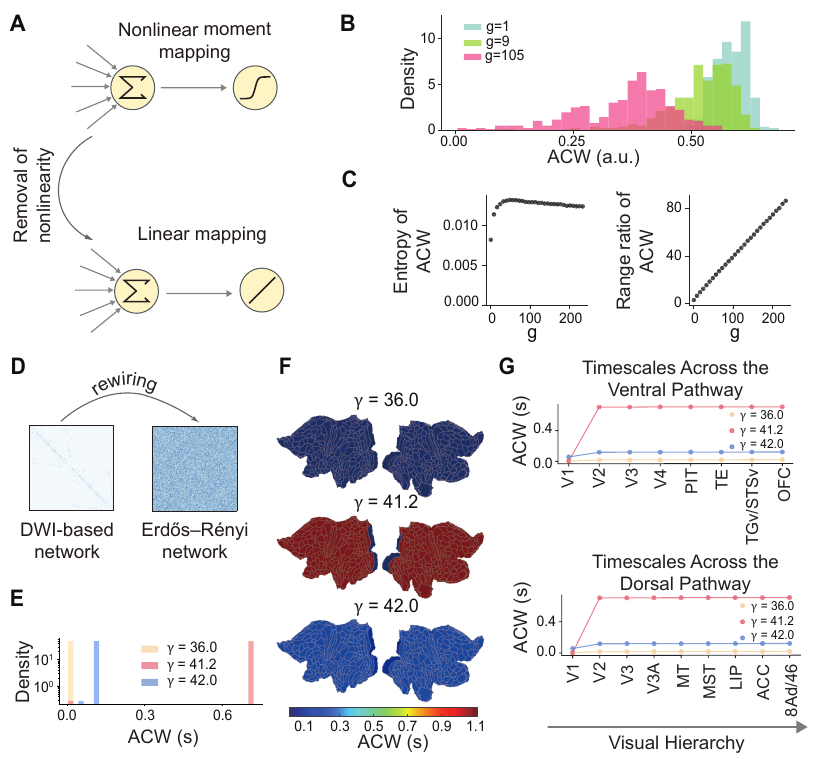}
   \caption{\textbf{Nonlinearity and structural connectome jointly shape timescale hierarchy.} (A) Schematic of a control experiment in which the nonlinear mean field model is replaced by a general linear process embedded within the DWI-informed structural connectome. (B) Distribution of intrinsic timescales across brain regions for different values of cross-regional coupling strength $g$. The distribution of intrinsic timescales broadens and becomes more diverse only when $g$ becomes extremely large ($g=105$). (C) Changes in the entropy and range ratio of intrinsic timescales as a function of $g$. (D) Schematic of the structural disruption experiment, where the DWI-based structural connectivity is randomly rewired into an Erdős–Rényi network. (E) Distribution of intrinsic timescales across brain regions under different dynamical regimes. (F) Intrinsic timescales mapped to the cortical surface under different regimes. (G) Variations in intrinsic timescales during visual signal processing along the ventral and dorsal pathways.\label{fig:compare MNN linear model}} 
\end{figure}

First, to investigate the role of biophysical nonlinearity in shaping the structure of intrinsic timescales, we consider a general linear process embedded with the DWI-informed structural connectome, as illustrated in Fig.~\ref{fig:compare MNN linear model}(A). This linear model has been employed in previous studies to describe the propagation of neural activity across a network over time \cite{Wang2019prl, zamora2016functional}. Its dynamic evolution is governed by the following equation
\begin{equation}
    \begin{aligned}
    \tau \frac{dx_i}{dt}=-x_i+g\sum_{j=1}^NA_{ij}(x_j-x_i)+I_{i,{\rm ext}}, \label{eq:linear diffusion model}
    \end{aligned}
\end{equation}
where $x_i$ represents the activity state of the $i$-th brain region, $I_{i,{\rm ext}}(t)$ is an external input, $A$ is the structural connectivity matrix determined by DWI data, $g$ represents the cross-regional coupling strength, and $\tau$ is a time constant in arbitrary unit (with $\tau$ set to 1 a.u. in this context). The matrix $A$ is normalized by the largest eigenvalue of the Laplacian matrix $H=D-A$ as described in \cite{Wang2019prl}, where $D$ is a diagonal matrix representing the weighted degrees of $A$. Since the Laplacian matrix $H$ is symmetric positive semi-definite, the eigenvalues of $-(I+gH)$ are negative for all $g>0$. Consequently, this linear system does not exhibit phase transitions as the parameter $g$ varies.

To characterize the temporal dynamics of the linear model during visual signal processing, we apply continuous stimuli with a mean of 0 and a standard deviation of 1 to V1 in both hemispheres and numerically simulate the responses of all brain regions, following the same procedure in the main text. The ACWs, the half-lives of autocorrelation functions (ACFs) of these responses, are then used to quantify the intrinsic timescales of different brain regions. Figure \ref{fig:compare MNN linear model}(B) shows the distribution of the intrinsic timescales across regions for different cross-regional coupling strengths $g$. When $g$ is relatively small ($g=1, 9$), the ACW is narrowly distributed with an approximately unimodal distribution. Only when $g$ becomes extremely large ($g = 105$) does the ACW distribution gradually broadens and becomes relatively more diverse. To further quantify the diversity and disparity in intrinsic timescales across regions, we compute the entropy and range ratio of the ACW across different dynamical regimes. Figure \ref{fig:compare MNN linear model}(C) illustrates changes in these metrics as cross-regional coupling strength $g$ varies. For the linear model, the ACW range ratio increases linearly with $g$, while the entropy gradually saturates as $g$ grows. This suggests that in the absence of biophysical nonlinearity, achieving more diverse and disparate temporal dynamics requires substantially larger values of $g$. In contrast, the biophysical nonlinearity incorporated into our mean field model is able to enhance the variability and heterogeneity of intrinsic timescales by fine-tuning parameters close to the critical point, without requiring extreme parameter values. Thus, biophysical nonlinearity is an essential ingredient for the human brain to achieve more diverse temporal dynamics through criticality.

Second, to explore the role of the structural connectome in shaping the intrinsic timescales of brain regions, we randomly rewired the DWI-based structural connectivity into an Erd\H{o}s–Rényi network, as illustrated in Fig.~\ref{fig:compare MNN linear model}(D). The surrogate network preserves the same connection density as the original structural connectome, with connection weights sampled uniformly from the range [0, 1). We then perform linear stability analysis to determine the dynamical regimes of the mean field model on the surrogate network, revealing that the model’s critical point shifts to $\gamma = 41.2$. 
To examine changes in timescales with the surrogate network across different regimes, we follow the same procedure as in the main text to estimate ACW during visual signal processing. Figure \ref{fig:compare MNN linear model}(E) illustrates the distribution of intrinsic timescales across brain regions under subcritical ($\gamma=36.0$), critical ($\gamma=41.2$), and supercritical ($\gamma=42.0$) dynamical regimes. Notably, the ACWs in all these regimes remain narrowly distributed. We further examine the spatial gradient of ACWs by mapping them to the cortical surface, as illustrated in Fig.~\ref{fig:compare MNN linear model}(F). The results reveal a homogeneous distribution of ACWs across brain regions in all these regimes, with no distinct hierarchical organization along the visual processing pathways. Specifically, as shown in Fig.~\ref{fig:compare MNN linear model}(G), the ACW rapidly saturates as soon as it enters adjacent sensory region (V2) of the stimulation site (V1) along both the ventral and dorsal pathways. 

Together, these findings suggest that the structural connectome provides a scaffold for the timescale hierarchy across brain regions, while criticality serves as a mechanism to further amplify the diversity and disparity of temporal dynamics. This result highlight the complex interactions between structure and dynamics in shaping the hierarchical signal processing in the brain.

 \section{Validating the computational models against neural imaging data}

\begin{figure}
    \centering
\includegraphics[width=1\textwidth]{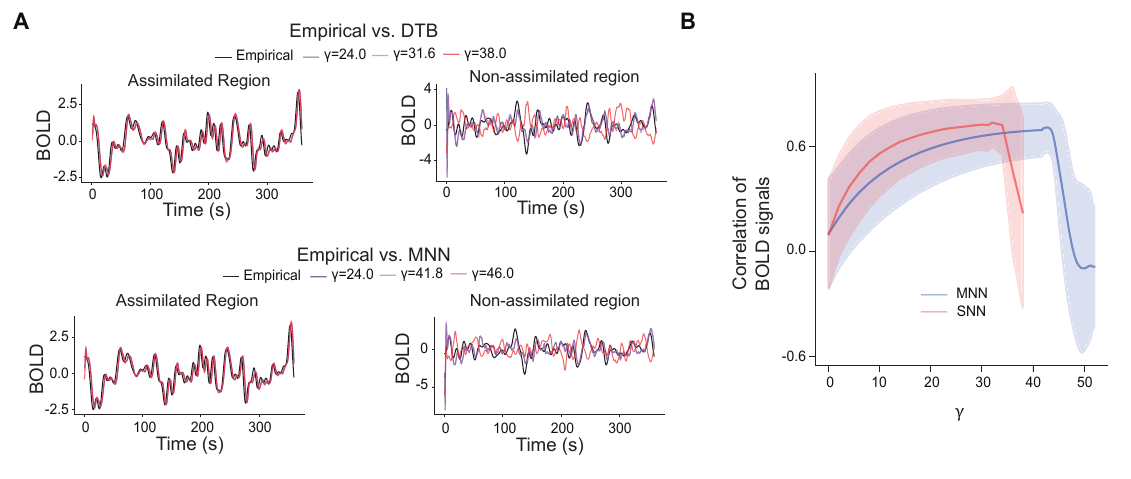}
   \caption{\textbf{Validating the computational models against neural imaging data}. (A) Time series of simulated BOLD signals from the DTB brain (upper panels) and the mean field model (lower panels) compared to data of the real brain. (B) Correlation coefficients between simulated and experimental BOLD signals for varying cross-regional coupling strengths. Solid curves and shades indicate the average and std across brain regions. Simulations of the DTB were run with an average synaptic in-degree of 500 and 9.45 million neurons. \label{fig:bold_compare_dtb_MNN}} 
\end{figure}

Next, we examine how well the mean field model and the DTB are able to reproduce experimental measurements of the real brain. For this purpose, we first compare the BOLD signals obtained from fMRI and the simulated BOLD signals using the mean field model and the DTB model. The upper row of Fig.~\ref{fig:bold_compare_dtb_MNN}(A) show typical time series of the simulated BOLD signals of the DTB brain as compared to that of the real brain, for different values of cross-regional coupling strengths. We find that for the assimilated brain region [top-left panel of Fig.~\ref{fig:bold_compare_dtb_MNN}(A)], the simulated BOLD signals well agree with those measured from the real brain. For the non-assimilated brain regions, this agreement is limited to the asynchronous activity ($\gamma=24$) and activity near the critical regime ($\gamma=31.6$). For stronger cross-regional coupling, the simulated BOLD signals no longer track the real data ($\gamma=38$) [top-right panel of Fig.~\ref{fig:bold_compare_dtb_MNN}(A)]. Qualitatively similar results are found in the simulated BOLD signals generated from the mean field model [lower row of Fig.~\ref{fig:bold_compare_dtb_MNN}(A)].

To quantify the similarity between the simulated and experimental BOLD signals, we calculate their correlation coefficients for each brain region. As shown in Fig.~\ref{fig:bold_compare_dtb_MNN}(B), both the mean field model and the DTB model display weak correlation in BOLD signals with the real brain when the cross-regional coupling is weak. This is expected as neural firing activity is dominated by local cortical circuits without significant contributions from whole brain connectome nor assimilated external inputs. As the cross-regional coupling strength increases, the correlation in BOLD signal also increases and reaches a peak of $\rho=0.73$ at $\gamma=32$ for the DTB model and $\rho=0.71$ at $\gamma=42.8$ for the mean field model. As cross-regional coupling further increases ($\gamma=38$ for the DTB and $\gamma=46$ for the mean field model), a negative correlation between simulated and experimental BOLD signals emerges (also evident in the right panel of Fig.~\ref{fig:bold_compare_dtb_MNN}(A)), reducing the similarity between the model and the real human brain.

We find that the increases in correlation between simulated and experimental BOLD signals within the asynchronous regime are caused by increases in signal-to-noise ratio. That is, as the cross-regional coupling becomes stronger, which are excitatory only, the influence from assimilated brain region on other brain areas becomes stronger. As a result, the simulated BOLD signal becomes less noisy, resulting in an overall increase in correlation coefficient. 

\section{Data acquisition and preprocessing}
Biological data used in this work were collected previously and were reported in our previous works~\cite{Lu2024nsr,Lu2024nat_cs}. We scanned multimodal MRI from a single subject, via a 3 Tesla MR scanner. A high-resolution T1-weighted (T1w) image acquired with a rapid gradient echo sequence, as well as multi-shell DWI and fMRI data acquired with gradient echo-planar imaging (EPI) sequences, were obtained to extract the VBM of gray matter, structural connectivity and BOLD signals, respectively. After preprocessing, a series of data-cleaning procedures were implemented to integrate multimodal neuroimaging data into our DTB model more effectively, resulting in a cortico-subcortical model with a total of 56,493 voxels.

\end{document}